\documentclass[a4paper]{aa}
\usepackage{graphicx}
\usepackage{natbib} 

\bibpunct{(}{)};{a}{}{,} 

\def\O{{\cal O}}
\def\C{{\cal C}}
\def\text#1{{\rm #1}}
\def\qaq{{\quad\text{and}\quad}}
\def\implies{{\quad\Longrightarrow\quad}}

\def\Rate{\mathit{\Gamma}}
\def\vf{\vec{f}}

\def\freq{\omega}

\def\n{{\rm n}}
\def\p{{\rm p}}
\def\e{{\rm e}}
\def\c{{\rm c}}
\def\s{{\rm s}}
\def\mb{m_\mathrm{b}}
\def\eps{\varepsilon}

\def\X{{\rm X}}
\def\Y{{\rm Y}}

\def\d{\delta}
\def\D{\Delta}

\def\vnabla{\vec{\nabla}}
\def\vv{\vec{v}}
\def\Dv{\Delta}
\def\vDv{\vec{\Dv}}
\def\vx{\vec{x}}

\def\S{{\cal S}}
\def\E{{\cal E}}
\def\mut{\widetilde{\mu}}
\def\entr{\alpha}
\def\K{{\cal K}}

\def\vR{\vec{R}}
\def\vS{\vec{S}}
\def\vT{\vec{T}}

\def\k{\kappa}
\def\kn{{\k_\n}}
\def\kc{{\k_\c}}
\def\Gn{{\gamma_\n}}
\def\Gc{{\gamma_\c}}
\def\GX{{\gamma_\X}}

\def\xc{x_\c}      
\def\xn{x_\n}      

\def\dbeta{\d\beta}

\def\nuc{\mathrm{nuc}}

\def\gcm3{ {\rm g\,cm}^{-3} }
\def\vnabla{\vec{\nabla}}
\def\c{{\rm c}}

\def\Msol{M_{\odot}}
\def\km{\mathrm{km}}

\def\ord{\mathrm{o}}
\def\sf{\mathrm{s}}

\def\stat{{(0)}}
\def\st{\widetilde{s}}

\newcommand{\eqq}[1]{(\ref{#1})}
\newcommand{\lp}{ \left(}
\newcommand{\rp}{ \right)}

\begin{document}

\title{Adiabatic oscillations of non--rotating superfluid neutron stars}

\author{Reinhard Prix\inst{1}
 \and Michel Rieutord\inst{2,3}}
\institute{Department of Mathematics, University of  Southampton, Southampton SO17 1BJ, UK
\and
Laboratoire d'Astrophysique de Toulouse, Obs. Midi-Pyr\'en\'ees, 14 avenue E.~Belin, 31400 Toulouse, France
\and Institut Universitaire de France}

\offprints{R.~Prix,\\ \email{r.prix@maths.soton.ac.uk}}

\date{Received ../ Accepted ..}
        
\abstract{We present results concerning the linear (radial and
  non--radial) oscillations of non--rotating superfluid neutron stars
  in Newtonian physics. We use a simple two--fluid model to describe
  the superfluid neutron star, where one
  fluid consists of the superfluid neutrons, while the second fluid
  contains all the comoving constituents (protons,
  electrons). The two fluids are assumed to be ``free'' in the sense
  of absence of  vortex--mediated forces like mutual friction or
  pinning, but they can be \emph{coupled} by the equation of
  state, in particular by entrainment. 
  We calculate numerically the eigen-frequencies and -modes of
  adiabatic oscillations, neglecting beta--reactions that
  would lead to dissipation.  We find a doubling of all
  acoustic--type modes (f--modes, p--modes), and confirm the
  absence of g--modes in these superfluid models.  We show
  analytically and numerically that \emph{only} in the case of
  non--stratified background models (i.e. with no composition
  gradient) can these doublets of acoustic modes be separated into
  two distinct families, which are characterised by either co-- or
  counter--moving fluids respectively, and which are sometimes
  referred to as ``ordinary''- and ``superfluid''  modes. 
  In the general, stratified case, however, this separation is
  \emph{not} possible, and these acoustic modes can not be classified
  as being either purely ``ordinary'' or ``superfluid''. 
  We show how the properties of the two--fluid modes change as
  functions of the coupling by entrainment. We find avoided
  mode-crossings for the stratified models, while the crossings  are 
  not avoided in the non--stratified, separable case.
  The oscillations of normal-fluid neutron stars are recovered
  as a special case simply by locking the two fluids together. 
  In this effective one--fluid case we find the usual singlet f--
  and p--modes, and we also find the expected g--modes of stratified
  neutron star models.
\keywords{Stars:neutron -- Stars:oscillations -- neutron stars: superfluidity}}

\maketitle

\section{Introduction}

The study of stellar oscillations has proved very fruitful in
improving our understanding of the inner structure and dynamics of
stars (the terms helio-- and astro--seismology have been
coined), for which the oscillation modes can often be observed rather
directly. The best developed example of this probing of the
internal structure of an astrophysical body via its oscillations is
probably the Earth. 
In the case of neutron stars, the observation of oscillations is
unfortunately not possible in such a direct way, and has not yet
been achieved. In practically all cases we can only observe the
regular radio--pulses of neutron stars, which are virtually unaffected 
by its oscillations and give information mostly about their rotation
rate. 
Nevertheless this field bears great potential interest: 
on one hand the better understanding of neutron star oscillations
could eventually  help to elucidate the phenomenon of glitches, which
is probably the most striking and puzzling aspect of observed neutron
star dynamics. This phenomenon still represents somewhat of a mystery,
even though the crucial role of superfluidity seems well established (see
\citet{link00:_probing_NS_glitches,carter00:_centr_buoyancy} for
recent discussions).  
On the other hand, several highly sensitive gravitational wave
detectors are expected to reach their full sensitivity within the next
few years, and neutron star oscillations represent one of the
potentially most interesting sources of gravitational waves.
Gravitational wave detection could open a new and complementary
observational window onto neutron stars, which would allow us to learn 
much more about their inner structure and dynamics than it is currently
possible with the purely electro-magnetic observations.

Currently most studies of neutron star oscillations are based on
simple perfect fluid models, which neglects the crucial importance of
superfluidity in neutron stars. The presence of substantial amounts
of superfluid matter in neutron stars is backed by a number of
theoretical calculations of the state of matter at these extreme
densities (e.g. see
\citet{baldo92:_superfl_neutr_star_matter,sjoberg76:_effect_mass}),  
and by the qualitative success of superfluid models to accommodate
observed features of glitches and their relaxation.

The first study to point out the importance of superfluidity for the
oscillation properties of neutron stars was by \citet{Epstein88}, who
has argued in a local (sound wave) analysis that superfluidity
should lead to new modes and modify the previously known modes. 
\citet{Lindblom94} have argued further for the existence of these modes,
but failed to find them numerically. \citet{Lee95} presented the first
numerical results indicating the presence of new modes that did not
exist in perfect fluid models, and the absence of g--modes which would
have been present in the non--superfluid case.
A local analysis by \citet{andersson01:_dyn_superfl_ns} has
given further analytic evidence for the absence of g--modes in simple
superfluid models. The relativistic numerical analysis by
\citet{comer99:_quasinorm_modes_GR_superfl_NS} has shown an
effective doubling of  acoustic modes in superfluid models with
respect to the normal fluid case. Recently this work has been
extended by \citet{andersson02:_oscil_GR_superfl_NS} to include
entrainment, and they have shown that avoided mode crossings occur
when one varies the entrainment parameter.  
The relevance of superfluid oscillations for gravitational wave
detection has been discussed by \citet{andersson01:_probing_ns_superfl}. 
Some studies have also started to look at oscillations of rotating
superfluid neutron stars \citep{Lindblom00} and \citet{Sedrakian00}. 

Despite the number of studies on oscillations of non--rotating
superfluid neutrons stars, we think that this problem still deserves
attention and that several points needed to be clarified.
In particular 
it is worth emphasising the importance of stratification for the
nature of superfluid oscillations, a point that has not yet been fully 
appreciated.
We  demonstrate here that only in non--stratified models can the
eigenmode spectrum be separated into two families of modes, one of
which is \emph{identical} to the case of a normal--fluid star, while
the other is characterised by counter--motion of the two fluids and
vanishing gravitational perturbation. These two distinct families are
usually referred to as ``ordinary'' and ``superfluid'' modes.
Stratification of the background star, however, couples these distinct  
mode-families and renders them non--separable. As a consequence 
every mode shares qualitative properties of both families to some
extent, and the resulting mode spectrum consists of modes that bear no
direct connection to the normal fluid case. 
The two--fluid model used here to describe superfluid neutron stars
is practically equivalent to those use in previous studies, and we refer
the reader to \citet{andersson01:_dyn_superfl_ns} and
\citet{prix02:_slowl_newton} for a more extensive discussion about its 
physical motivations and justification.

The plan of this paper is as follows: in
Sect.~\ref{sec:GenTwoFluidModel} we introduce the basic equations 
for the general two--fluid neutron star model, and in
Sect.~\ref{sec:LinPert} we develop its linear perturbation equations
and show how to recover the special case of a single perfect fluid.
In Sect.~\ref{sec:Adiabatic} we specialise to the simpler case of
adiabatic oscillations of free,  cold fluids, and we derive the
necessary boundary conditions. In this section we also show that the
separation into two distinct mode families is possible only for
non--stratified models. Sect.~\ref{sec:NumSol} presents the  
numerical results concerning the background models, the eigenmode
spectrum and its dependence on entrainment (resulting in avoided crossings),
as well as the one--fluid results where we recover the expected
composition g--modes. We present our conclusions and a discussion of
necessary future work in Sect.~\ref{sec:Conclusions}.

\section{The general two--fluid neutron star model}
\label{sec:GenTwoFluidModel}

\subsection{The general two--fluid equations}

The equations and notation for the Newtonian two--fluid neutron star
model used here are based on a more general 
formalism described in \citet{prix02:_variat}, which is the Newtonian
analogue of a generally relativistic framework developed by \citet{Carter89}.
In this section we will briefly summarise the general model and 
equations relevant for the present work, and we refer the reader to
\citet{prix02:_variat} for the derivation and more detailed discussion
of this model.

We describe a neutron star as a system consisting of two
fluids: a superfluid of neutrons, and a normal fluid of comoving
constituents, which include protons, electrons and entropy (and
generally further particles like muons etc). 
We denote the particle number densities for neutrons, protons and
electrons as $n_\n$, $n_\p$ and $n_\e$ respectively, and we use $s$
for the entropy density. The velocities of the two fluids are
$\vv_\n$ for the neutron fluid, and $\vv_\c$ for the fluid of comoving
constituents, the relative velocity $\vDv$ between the two fluids is
therefore 
\begin{equation}
  \label{eq:RelVel}
  \vDv \equiv \vv_\c - \vv_\n\,.
\end{equation}
On the local (``mesoscopic'') level, a superfluid is constrained to be
in a state of irrotational flow, and its angular momentum will be
accommodated by a lattice of ``microscopic'' vortices. 
However, for the global description of a neutron star we are more
interested in a ``macroscopic'' description of the superfluid,
obtained by averaging over volume elements containing a large number
of vortices, but which are small compared to the dimensions of the 
neutron star. In this macroscopic description, the superfluid has a
continuous vorticity and nearly behaves like an ordinary fluid (apart
from small anisotropies due to the vortex-tension, which we will
neglect). The absence of (local) viscosity still allows the superfluid
to move relative to the normal fluid, but the presence of the vortex
lattice now allows for a direct force between the two fluids. In the
case of this force being zero, the vortex lattice moves with the
superfluid and we refer to this situation as free fluids. If the
vortices are ``locked'' to the normal fluid (e.g. as can happen in the
crust), the mutual force will be non-zero but strictly
non--dissipative. This is known as ``vortex pinning''. Only in the
intermediate cases, where a friction force causes the vortices
to have a different velocity from both the normal fluid and the
superfluid, energy is dissipated. In this case the mutual force is
usually referred to as ``mutual friction''.  

An essential simplification of the present two--fluid model is that 
we neglect all electrodynamic effects, as we assume the charge
densities of protons and electrons to be strictly balanced, i.e.
\begin{equation}
  \label{eq:ChargeNeut}
n_\e = n_\p\,.
\end{equation}
Therefore we can effectively eliminate the electrons from our
description, as their density and velocity is entirely specified by
the protons. 

We note that in ``transfusive'' models (as first set up in
\citet{Langlois98}), i.e. models which allow for $\beta$--reactions
($\n\rightleftharpoons \p+\e + \bar{\nu}$)  between the two fluids,
the total mass in the reaction has to be conserved in a consistent
Newtonian description. Therefore we set  
\begin{equation}
  \label{eq:BaryonMass}
\mb \equiv m_\n \equiv m_\p + m_\e\,.  
\end{equation}
The respective mass densities of the two fluids can now be written as
\begin{equation}
  \label{eq:MassDensities}
  \rho_\n = \mb n_\n\,,\qaq 
  \rho_\c = \mb n_\p\,,
\end{equation}
and the total mass density $\rho$ is simply $\rho=\rho_\n + \rho_\c$.

The (local) \emph{kinematics} of the system is completely described
(up to arbitrary rotations and boosts) in terms of $\rho_\n$,
$\rho_\c$, $s$ and $\Dv^2$. The \emph{dynamics} is determined
by the internal energy density function $\E$ or equation of state,
which is a function of the form $\E=\E(s, \rho_\n, \rho_\c, \Dv^2)$.
This energy function defines the first law of thermodynamics for
this system by its total differential, namely
\begin{equation}
  \label{eq:FirstLaw}
  d\E = T \,d s + \mut^\c \,d \rho_\c + \mut^\n \,d \rho_\n +\entr
\,d\Dv^2\,.
\end{equation}
This differential defines the dynamic quantities, namely the
temperature $T$, the specific chemical potentials $\mut^\c$,
$\mut^\n$, and the entrainment $\entr$ as the thermodynamic conjugates
of the kinematic quantities $s$, $\rho_\c$, $\rho_\n$ and $\Dv$.
The specific chemical potentials $\mut$ are related to the more common
definition of the chemical potentials $\mu$ via  $\mut \equiv \mu/\mb$.
We note that $\mu^\c$ is not simply the proton chemical potential,
because adding a proton in this model implies adding an electron as
well (due to (\ref{eq:ChargeNeut})), and therefore one can see that
$\mu^\c=\mu^\p + \mu^\e$, where $\mu^\p$ and $\mu^\e$ are the
respective proton and electron chemical potentials.

The function $\entr$ defined in (\ref{eq:FirstLaw}) reflects  the
dependence of the internal energy on the relative velocity $\Dv$
between the two fluids, which characterises the so--called
entrainment effect. This entrainment function $\entr$ has
dimensions of a mass density, and it will be useful in the following
to define the two dimensionless entrainment functions $\eps_\n$ and
$\eps_\c$ as  
\begin{equation}
  \label{eq:EpsDef}
  \eps_\c \equiv {2\entr\over \rho_\c}\,, \qaq \eps_\n \equiv
  {2\entr\over \rho_\n} = {\rho_\c \over \rho_\n} \eps_\c\,.
\end{equation}
The ``generalised'' pressure $P$ is introduced as the Legendre-conjugate
of the energy density, namely by the usual thermodynamic relation  
\begin{equation}
  \label{eq:Pressure}
  \E + P = \rho_\n \mut^\n + \rho_\c \mut^\c + s T\,,
\end{equation}
which results in the total differential of the pressure function
$P(\mut^\c,\mut^\n,T,\Dv^2)$: 
\begin{equation}
  \label{eq:PressureVariation}
  d P = \rho_\n \, d\mut^\n + \rho_\n \, d\mut^\c + s\, d T - \entr\, d\Dv^2\,.
\end{equation}
This generalised pressure $P$ can be seen to reduce to the usual
definition of the pressure of a perfect fluid in the case of $\Dv=0$.

The equations of motion of this two--fluid system are derived from a
``convective'' variational principle in \citet{prix02:_variat}, and
here we only present the resulting equations. The conservation of
energy results in the following equation: 
\begin{equation}
T\Rate_\s = \vf\cdot\vDv + \beta \Rate_\n  \,,
\label{eq:Energy}
\end{equation}
where $\Rate_\s$ and $\Rate_\n$ are the creation rates of entropy and
neutrons respectively, i.e. 

\begin{equation}
\Rate_\s \equiv \partial_t s + \vnabla\cdot(s \vv_\c)\,,\;\;\text{and}\;\;
\Rate_\n \equiv \partial_t \rho_\n + \vnabla\cdot(\rho_\n \vv_\n)\,,
\end{equation}
while the proton creation rate $\Rate_\c$ has to satisfy
$\Rate_\c=-\Rate_\n$ for baryon conservation.
The quantity $\beta$ in (\ref{eq:Energy}) characterises the deviation
from chemical equilibrium and its explicit expression if found as

\begin{equation}
  \label{eq:ChemicalEquilibrium}
  \beta \equiv 
  \mut^\c - \mut^\n - {1\over2}\left(1 - 2\eps_\n\right)\,{\Dv^2}\,.
\end{equation}
This is the Newtonian analogue of a result first found in the
 relativistic transfusive model by \citet{Langlois98}.
We note that there is an additional kinetic term with respect to
the naive $\mut^\c - \mut^\n$ in the case of relative motion.
The mutual force density $\vf$ (the sign convention is such that
this force acts \emph{on} the neutron fluid) is a
direct interaction force between the two fluids, which in the case of
a superfluid stems from vortex interactions like pinning or
mutual friction.   In order to ensure explicitly that
the second law of thermodynamics, i.e. \mbox{$\Rate_\s\ge0$}, is
satisfied  by (\ref{eq:Energy}), we can write the neutron creation rate
$\Rate_\n$ and the mutual force $\vf$ in the form 
\begin{eqnarray}
\Rate_\n &=& \Xi \beta\,, \hspace{1.7cm}\quad\text{with} \quad \Xi\ge0 \,,
\label{eq:NCreation}\\
\vf &=& \eta \vDv + \vec{\kappa}\times \vDv \,, \quad\text{with}
\quad \eta \ge 0\,, \label{eq:MutualForce}
\end{eqnarray}
where the non--negative functions $\Xi$ and $\eta$ govern the 
beta--reaction rate and the friction force, while the vector
$\vec{\kappa}$ allows for a non--dissipative Magnus--type force
(i.e. orthogonal to the relative motion).
A non--transfusive model, i.e. one that does not allow for beta
reactions  $\n\rightleftharpoons \c$, has $\Xi=0$, and free
vortices correspond to $\eta=0$ and  $\vec{\kappa}=0$, while
pinned vortices correspond to $\eta=0$ and $\vec{\kappa}\not=0$.

The momentum equation for the superfluid neutrons is given by
\begin{equation}
  (\partial_t + \vv_\n\cdot\vnabla)(\vv_\n + \eps_\n \vDv) +
  \vnabla(\mut^\n + \Phi) + \eps_\n \Dv_a\vnabla v_\n^a =
  {1\over\rho_\n}\vf,\quad
  \label{eq:EOMn}
\end{equation}
while the equation for the normal fluid reads as
\begin{eqnarray}
(\partial_t + \vv_\c\cdot\vnabla)( \vv_\c - \eps_\c \vDv) +
  \vnabla(\mut^\c +\Phi) - \eps_\c \Dv_a \vnabla v_\c^a \quad & &\quad\nonumber \\
+ {s\over\rho_\c}\nabla T = -{1\over \rho_\c} \vf + \left(1-\eps_\c - \eps_\n \right)
  {\Rate_\n \over \rho_\c} \vDv\,. & &
\label{eq:EOMc}
\end{eqnarray}
The gravitational potential $\Phi$ is related to the mass densities
$\rho_\n$ and $\rho_\c$ via the Poisson equation  
\begin{equation}
\nabla^2 \Phi = 4\pi G (\rho_\n + \rho_\c)\,.
\label{eq:Poisson}
\end{equation}

\subsection{The static equilibrium background}
\label{sec:StaticBg}
We consider a static background star, so we set
\begin{equation}
\vv_\n=\vv_\p=\vDv=0\,, \qaq \Rate_\n=\Rate_\c=\Rate_\s =0\,,
\label{eq:StaticBg}
\end{equation}
which by (\ref{eq:MutualForce}) implies the vanishing of the mutual
force, i.e. $\vf = 0$. Because chemical reactions would not be
negligible on long timescales, we also assume the background star to
be in chemical equilibrium, i.e.
\begin{equation}
  \label{eq:BgChemEquil}
\beta = \mut^\c - \mut^\n = 0\,. 
\end{equation}
These equilibrium conditions reduce the equation of motion
(\ref{eq:EOMn}) to 
\begin{equation}
\vnabla \mut^\n = \vnabla \mut^\c = - \vnabla \Phi \,,
\label{eq:MuEqu}
\end{equation}
and with (\ref{eq:EOMc}) this also implies that the background star is
in thermal equilibrium, i.e. \mbox{$\vnabla T=0$}.

The static background has to be spherically symmetric, and therefore
(\ref{eq:MuEqu}) and (\ref{eq:Poisson}) lead to the following equation
for the background:
\begin{equation}
  \label{eq:BgConf}
  \mut''(r) + {2\over r} \mut'(r) = - 4\pi G \rho(r)\,,
\end{equation}
where we have introduced the equilibrium chemical potential
$\mut\equiv \mut^\n = \mut^\c$, and the prime ($'$) denotes the radial
derivative $d/dr$.

The equation of state allows one to relate the equilibrium chemical
potential $\mut$ directly to the total mass density $\rho$ at constant
temperature, and therefore the background is fully determined.
The numerical method for solving this equation will be discussed in
sect.~\ref{sec:NumSol}. 

In the following it will be convenient to use the radius $R$ and
central density $\rho_0$ of the static background  as
basis units for length and mass density, so the corresponding
``natural unit'' for frequencies is $\sqrt{4\pi G \rho_0}$. All
equations in the following  are expressed in these natural
units except otherwise stated.

\subsection{Entrainment and effective masses}
\label{sec:Entrainment}

For small relative velocities $\Dv$, we can separate the ``bulk''
equation of state from the entrainment by expanding $\E(s, \rho_\n,
\rho_\c; \Dv^2)$ in terms of $\Dv$, i.e. by writing
\begin{equation}
  \label{eq:SmallDvExpansion}
  \E = \E^\stat(s, \rho_\n, \rho_c) +
  \entr^\stat(s, \rho_\n,\rho_\c)\Dv^2 + \O(\Dv^4)\,,
\end{equation}
where the quantities $\E^\stat$ and $\entr^\stat$ are evaluated at
zero relative velocity $\Dv$.
The background equation of state  $\E^\stat$  is therefore decoupled
from the entrainment function $\entr^\stat$, and we can  specify these  
two functions independently. 
The link between the entrainment function $\entr$ and the equivalent
description in terms of effective masses $m^*$ \citep{Andreev75}
has been discussed in previous work \citep{prix02:_slowl_newton},
and it can be shown that one can express $\entr$ in terms of the
proton effective mass $m_\p^*$ (which is generally a function of the
densities), in the form
\begin{equation}
  \label{eq:EffectiveMassEntr}
  2\entr = \rho_\c \left( 1 - {m_\p^*\over\mb}\right)\,.
\end{equation}
The dimensionless entrainment functions\footnote{
We note that \citet{Lindblom00} and more recently
\citet{andersson02:_oscil_GR_superfl_NS} have used a slightly
different dimensionless function  $\epsilon$ to characterise the
entrainment effect. The relation between $\epsilon$ and $\eps_c$ is
given by  $ \epsilon = \eps_c \rho_\c / (\rho_\n - \eps_c \rho)$.}
 $\eps_\n$ and $\eps_\c$ can
then be expressed according to (\ref{eq:EpsDef}) as
\begin{equation}
  \label{eq:EffectiveMassEps}
  \eps_\c = 1 - {m_\p^* \over \mb}\,,\qaq
  \eps_\n = {\rho_\c\over \rho_\n} \, \eps_\c\,.
\end{equation}

\section{Linearised perturbation equations}
\label{sec:LinPert}

\subsection{Oscillations of superfluid neutron stars}
We consider small perturbations with respect to the static equilibrium
background described in Sect.~\ref{sec:StaticBg}. Linearising the
equations of motion (\ref{eq:EOMn}) and (\ref{eq:EOMc}) yields
\begin{eqnarray}
  \label{eq:LinearEOMn}
\rho_\n \partial_t ( \d \vv_\n + \eps_\n \,\d\vDv )+
\rho_\n \vnabla(\d\mut^\n + \d\Phi) =\d\vf\,,\qquad\qquad\quad\mbox{}\\ 
\rho_\c \partial_t( \d \vv_\c - \eps_\c \,\d\vDv) +
\rho_\c \vnabla(\d\mut^\c + \d\Phi) =-\d\vf -s \vnabla\d T,\quad\mbox{}
  \label{eq:LinearEOMc}
\end{eqnarray}
where $\d Q$ denotes the \emph{Eulerian} perturbation of the
quantity $Q$. The perturbation of the mutual force
(\ref{eq:MutualForce}) is given by
\begin{equation}
  \label{eq:df}
  \d\vf = \eta \d\vDv + \vec{\kappa}\times \d\vDv\,,
\end{equation}
and the linearised energy conservation (\ref{eq:Energy}) and
(\ref{eq:NCreation}) together with the condition of baryon
conservation lead to  
\begin{eqnarray}
  \label{eq:LinearCons}
  \partial_t\, \d s + \vnabla\cdot(s \, \d\vv_\c) &=& 0 \,,\\
  \partial_t\, \d \rho_\n + \vnabla\cdot(\rho_\n \,\d\vv_\n) &=& \;\;\;\Xi
  \,(\d\mut^\c - \d\mut^\n)\,,\\ 
  \partial_t\, \d \rho_\c + \vnabla\cdot(\rho_\c \,\d\vv_\c) &=& -\Xi
  \,(\d\mut^\c - \d\mut^\n)\,.
\end{eqnarray}
The perturbed Poisson equation (\ref{eq:Poisson}) reads (in
natural units) as
\begin{equation}
  \label{eq:LinearPoisson}
  \nabla^2 \d\Phi = \d \rho_\n + \d \rho_\c \,.
\end{equation}
The system is closed by specification of the mutual force
functions $\eta$ and $\vec{\kappa}$, the transfusion function
$\Xi$, and an equation of state 
which allows to express the dynamical quantities $\d T$,
$\d\mut^\n$ and  $\d\mut^\c$ in terms of the kinematic variables 
$\d s$, $\d\rho_\n$ and $\d\rho_\c$, thereby reducing the number of
unknown perturbation quantities to $13$, which corresponds exactly to
the number of equations.  

\subsection{The special case of normal--fluid neutron stars}
\label{sec:SingleFluidOsc}

It is interesting to compare the superfluid neutron star case with the
normal fluid case, where the  two constituents $n$ and $\c$ are
moving together and  form a single perfect fluid. This case is
obviously just a subclass of the two--fluid case discussed so far,
namely subject to the additional constraint $\vv_\n = \vv_\c$,
and therefore\footnote{\citet{Lindblom94,Lindblom95} have imposed $\d\beta=0$
  to recover the perfect fluid case. However, adiabatic
  oscillations of a perfect fluid only satisfy this condition in 
  \emph{non--stratified} stars (cf. sect.~\ref{sec:Num1fEigenmodes}),
  the constraint $\dbeta=0$ is therefore generally not met.}   
\begin{equation}
  \label{eq:1FluidLimit}
  \d\vv_\n  = \d\vv_\c = \d\vv\,, \quad\implies \d\vDv = 0\,.
\end{equation}
By linking the two constituents together, the degrees of freedom have
been reduced by three, and instead of the individual momentum
equations (\ref{eq:LinearEOMn}) and (\ref{eq:LinearEOMc}), now only
the sum of momenta can be required to be conserved, i.e. 
\begin{equation}
  \label{eq:1FluidEOM}
  \rho \,\left( \partial_t \d\vv + \vnabla \d\Phi\right)  + 
\rho_\n \vnabla \d\mut^\n + \rho_\c \vnabla \d\mut^\c + s\vnabla \d T
= 0 \,.
\end{equation}
We introduce the notation
\begin{equation}
  \rho_\c = \xc \rho\,,\quad
  \rho_\n = \xn \rho\,,\qaq
  s = \st \rho\,,
\end{equation}
for the proton and neutron fractions $\xc$ and $\xn$, and the 
specific entropy $\st$, which allows us to rewrite the one--fluid
equation of motion (\ref{eq:1FluidEOM}) in the slightly more familiar
form   
\begin{equation}
  \label{eq:1FluidEuler}
  \partial_t \d\vv + \vnabla\left( {\d P \over \rho} + \d\Phi\right) - 
  \left[ \d\beta\, \vnabla \xc + \d T \, \vnabla \st \right] = 0\,,
\end{equation}
where we used the fact that the total pressure differential $d P$ of
(\ref{eq:PressureVariation})  in this perfect fluid case reduces to 
\begin{equation}
  \label{eq:dPressure1f}
  d P = \rho_\n \, d\mut^\n + \rho_\n \, d\mut^\c + s\, d T\,.
\end{equation}
We can now compare (\ref{eq:1FluidEOM}) to the standard expression for
the Euler equation of stellar oscillations of non--rotating stars
\citep{cox80:_theor_stell_pulsat,unno89:_nonrad_oscil_stars}, which is
usually written as
\begin{equation}
  \label{eq:StdOscillationEqs}
  \partial_t \d\vv + \vnabla\left( {\d P \over \rho} + \d\Phi\right)  
  + \C(r)\, (\vnabla\cdot\d\vv)\,\vec{A} = 0\,,
\end{equation}
where $\C(r)$ is a function of the background, and $\vec{A}$ is the
so--called Schwarzschild discriminant that is responsible for presence
of g--modes. By comparing eqs.~(\ref{eq:1FluidEOM}) and
(\ref{eq:1FluidEuler}), we see that $\vec{A}$ will be non--zero
(indicating the presence of g--modes) whenever $\vnabla \st\not=0$ or
$\vnabla\xc\not=0$. This reflects the well--known fact that any type
of stratification, either in specific entropy $\st$ or in the
chemical composition $\xc$, will result in g--modes, as pointed out 
by \citet{reisenegger92}.

\section{Adiabatic oscillations of "free", cold fluids}
\label{sec:Adiabatic}

\subsection{Reduction to a 1D eigenvalue problem}

In order to close the system of perturbation equations
(\ref{eq:LinearEOMn})--(\ref{eq:LinearPoisson}), we need specific
models for the mutual force $\vf$, the transfusion $\Xi$, in addition
to an equation of state of the form $\E(\rho_n,\rho_\c,\s)$, all of
which are highly dependent on microphysical models and are 
rather poorly known at the present stage. For this reason, we will 
postpone the inclusion of these effects to future work, and focus 
on the case of purely \emph{adiabatic} oscillations (i.e. $\Xi=0$) of
free fluids (meaning  $\vf=0$). We will further neglect
temperature effects  (which is generally a very good approximation
except for very young neutron stars), so we set $s=0$
and $T=0$. The resulting simplified system of 
equations is 
\begin{eqnarray}
  \partial_t \d \rho_\n + \vnabla\cdot\left(\rho_\n \,\d\vv_\n\right) &=& 0\,,
  \label{equ20} \\
  \partial_t \d \rho_\c + \vnabla\cdot\left(\rho_\c \,\d\vv_\c\right) &=& 0\,,\\
  \partial_t \left[ (1-\eps_\n) \,\d\vv_\n + \eps_\n \d\vv_\c \right] &=&
 - \vnabla\left(\d\mut^\n + \d\Phi\right)\,,\\
  \partial_t \left[ (1-\eps_\c)\, \d\vv_\c + \eps_\c \d\vv_\n \right] &=&
 - \vnabla\left(\d\mut^\c + \d\Phi\right)\,, \\
  \nabla^2 \d\Phi &=& \d \rho_\n + \d \rho_\c,. \label{equ24}
\end{eqnarray}
We point out that this system of equations is identical to the one
used in \citet{andersson01:_dyn_superfl_ns}, which was obtained from a
Newtonian limit of the relativistic equations. It is also related to
the equations of \citet{Lindblom94}, which are expressed in the
``orthodox'' formulation of superfluids
\citep{landau59:_fluid_mechanics}, while the present description is 
based on the ``canonical'' approach introduced by \citet{Carter89}.

Using the equation of state we can link the density perturbations
$\d\rho_\X$ to $\d\mut^\X$ (with the constituent index notation
$\X=\n,\,\c$)  to linear order, namely
\begin{equation}
  \label{eq:drhodmu}
  \begin{array}{r l}
  \d\rho_\n =& \S_{\n\n} \, \d\mut^\n + \S_{\n\c}\, \d\mut^\c\,,\\
  \d\rho_\c =& \S_{\c\n} \, \d\mut^\n + \S_{\c\c}\, \d\mut^\c\,,\\
  \end{array}
\end{equation}
where the symmetric ``structure matrix'' $\S_{\X\Y}$ is defined as
\begin{equation}
  \label{eq:DefSXY}
  \left(\S^{-1}\right)_{\X\Y} \equiv {\partial \mut^\X \over
  \partial \rho_\Y} = {\partial^2 \E \over \partial
  \rho_\X\, \partial \rho_\Y}\,,\qquad \X,\Y \in \{\n,\c\}\,.
\end{equation}
Due to the dual role of the pressure $P$ in (\ref{eq:Pressure})
and (\ref{eq:PressureVariation}), we can equivalently express
$\S_{\X\Y}$ as 
\begin{equation}
  \label{eq:SXYAlt}
  \S_{\X\Y} = {\partial \rho_\X \over \partial \mut^\Y} = 
{\partial^2 P \over \partial \mut^\X \,\partial \mut^\Y}\,.
\end{equation}
We note that although we have assumed free fluids, i.e. there is
no direct force acting between them, the fluids are nevertheless
locally coupled by the equation of state; we can distinguish two
sources of this coupling, one is due to the non--diagonal term
$\S_{\n\c}$ in (\ref{eq:drhodmu}), while the second is due to the
entrainment terms $\eps_\X$.

The background quantities $\mut^\X$ can be seen in
(\ref{eq:MuEqu}) to behave like the gravitational potential $\Phi$;
this means in particular that their gradient is always finite, even at
the surface.
Therefore $\d\mut^\X$ is finite everywhere, contrary to
$\d\rho_\X$ which can diverge at the surface when
$\rho'\rightarrow-\infty$.
This is seen from the relation  $\D\rho = \d\rho +
\xi^r\,\rho'(r)$ between the Lagrangian perturbation $\D\rho$ and the
Eulerian $\d\rho$, for a radial displacement $\xi^r$. 
On physical grounds $\D\rho$ must be bounded everywhere (as it
reflects the physical property of a fluid element), while the
first--order Eulerian quantity $\d\rho$ diverges at the surface
whenever $\rho'\rightarrow-\infty$ and $\xi^r\not=0$ at $r=R$. 
This might seem problematic for the validity of the equations,
but it only reflects the fact that in this case even an infinitesimal
displacement of the surface will lead to a \emph{finite} (as opposed
to infinitesimal) Eulerian density change there. By considering
Lagrangian instead of Eulerian variables, it can be shown that the
physical solution is still well behaved even if
$\d\rho\rightarrow\infty$ at the surface.   
In this case the first--order quantity $\d\rho$ no longer approximates
the physical Eulerian density change, but the divergence is such that
the \emph{Lagrangian} first--order quantity $\D\rho$ is still
perfectly regular. 
If one wanted to impose that $\d\rho$ should be bounded everywhere (as
did \citet{Lindblom94}), then this situation would be inverted and the
Lagrangian quantity $\D\rho$ would diverge, which is unphysical indeed.    

From a numerical point of view it seemed better to solve directly for the
well-behaved $\d\mut^\X$ instead of the potentially diverging
$\d\rho_\X$, by using (\ref{eq:drhodmu}) to substitute for
$\d\rho_\X$. We note that the coefficients $\S_{\X\Y}$ in this
expression will generally diverge (or vanish) at the surface,
depending on the equation of state and reflecting the behaviour of
$\d\rho_\X$.
The system of equations (\ref{equ20}-\ref{equ24}) for eigenmode
solutions of the form $\d Q(\vx,t) = \d Q(\vx)\,e^{i\freq t}$ now yields 
\begin{eqnarray}
  \vnabla\cdot( \rho_\n \d\vv_\n) &=& - i\freq  \left[ \S_{\n\n} \d\mut^\n +
    \S_{\n\c} \d\mut^\c \right]\,,   \label{eq:EigenvalueSystem2Da}\\
  \vnabla\cdot( \rho_\c \d\vv_\c) &=& - i\freq  \left[ \S_{\c\n} \d\mut^\n +
    \S_{\c\c} \d\mut^\c \right]\,,\label{mass2}\\
  \vnabla( \d\mut^\n + \d\Phi ) &=& - i\freq  \left[ (1-\eps_\n)\, \d\vv_\n +
  \eps_\c \, \d\vv_\c \right]\,,\\
  \vnabla( \d\mut^\c + \d\Phi ) &=& - i\freq  \left[ (1-\eps_\c)\, \d\vv_\c +
    \eps_\n \, \d\vv_\n \right]\,,\\
  \nabla^2 \d\Phi &=&  k_\n \, \d\mut^\n  + k_\c\, \d\mut^\c\,,
  \label{eq:EigenvalueSystem2Dz}
\end{eqnarray}
where we have introduced the convenient ``structure vector'' $k_\X$,
which is defined as 
\begin{equation}
  \label{eq:DefkX}
  k_\X \equiv \sum_{\Y=\n,\c} S_{\X\Y}\,,\qquad \X\in\{\n,\,\c\}\,.
\end{equation}
For a spherically symmetric background, we can separate the radial and
angular dependence and obtain solutions with definite quantum
numbers $l$ and $|m|\le l$ using the ansatz
\begin{eqnarray}
\d\Phi(r,\theta,\varphi) &=& \d\Phi(r) \,Y_l^m(\theta,\varphi)\,, \label{eq:HarmScalar}\\
\d\mut^\X(r,\theta,\varphi) &=& \d\mut^\X(r) \,Y_l^m(\theta,\varphi)\,,\\
\d\vv_\X(r,\theta,\varphi) &=& {W_\X(r)\over r}\vR_l^m +
V_\X(r)\vS_l^m - i U_\X(r)\vT_l^m, \label{eq:HarmVect}   
\end{eqnarray}
where the spherical harmonics $Y_l^m(\theta,\varphi)$ are the eigenfunctions
of $r^2 \nabla^2 Y_l^m = - l(l+1)\, Y_l^m$,
and $\vR$, $\vS$ and $\vT$ form the  orthogonal harmonic basis, defined as
\begin{equation}
  \label{eq:HarmonicBasis}
  \vR_l^m \equiv Y_l^m \vnabla r\,,\quad   
  \vS_l^m \equiv \vnabla Y_l^m \,,\quad
  \vT_l^m \equiv \vnabla\times \vR\,, 
\end{equation}
see \cite{rieu87} for details.
The three--dimensional eigenvalue problem
(\ref{eq:EigenvalueSystem2Da})-(\ref{eq:EigenvalueSystem2Dz}) has now
been reduced to the following one--dimensional problem:
\begin{eqnarray}
\left(r \rho_\n W_\n\right)' - {l(l+1)} \rho_\n V_\n &=& 
-i\freq r^2\left[\S_{\n\n} \d\mut^\n + \S_{\n\c}
\d\mut^\c\right], \label{eq:dmun}\\ 
\left(r \rho_\c W_\c\right)' - {l(l+1)}\rho_\c V_\c &=& 
-i\freq r^2\left[\S_{\n\c} \d\mut^\n + \S_{\c\c}
\d\mut^\c \right], \label{eq:dmuc}\\ 
-r\left({\d\mut^\n}' + \d\Phi'\right) &=& i\freq \left[
(1-\eps_\n) W_\n + \eps_\n W_\c) \right],\qquad\mbox{}\label{eq:Wn} \\
-r\left({\d\mut^\c}' + \d\Phi'\right) &=& i\freq \left[
(1-\eps_\c) W_\c + \eps_\c W_\n) \right], \label{eq:Wc}\\
-\d\mut^\n - \d\Phi &=& i\freq \left[
(1-\eps_\n) V_\n + \eps_\n V_\c)\right], \label{eq:Vn}\\
-\d\mut^\c - \d\Phi &=& i\freq \left[
(1-\eps_\c) V_\c + \eps_\c V_\n)\right], \label{eq:Vc}\\
\left(r^2\d\Phi'\right)' - {l(l+1)}\,\d\Phi   
&=& r^2\,(k_\n \,\d\mut^\n +  k_\c \,\d\mut^\c)\,. \label{eq:dPhi}
\end{eqnarray}
The axial velocity component $U_\X$ is decoupled and
corresponds to zero frequency $\freq=0$, therefore all non--zero
frequency eigenmodes are purely polar. We note that the horizontal
velocity equations (\ref{eq:Vn}) and (\ref{eq:Vc}) only hold for
$l>0$, because $\vec{S}_0^0 = 0$.

\subsection{Boundary and regularity conditions}
\label{sec:BoundaryConditions}

\subsubsection{At the center}
It can be shown that the representation
(\ref{eq:HarmScalar})-(\ref{eq:HarmVect}) of a regular physical
quantity requires the following asymptotic behaviour of the radial
functions as $r \rightarrow 0$,
\begin{equation}
  \label{eq:BCcentre}
  \d\mut^\X \sim \d\Phi \sim W_\X\sim V_\X \sim \O(r^l)\,,\qquad
\X \in \{\n,\,\c\}\,,
\end{equation}
where $\O(r^l)$ means of order $r^l$ \emph{or higher}. Another
requirement is the regularity of the solution at singular points of
the equations. From eqs.~(\ref{eq:Wn}) and (\ref{eq:Wc}) we see that
$W_\X$ must vanish as $\O(r)$ at the origin, while from
eqs. (\ref{eq:Vn}) and (\ref{eq:Vc}) we can derive the further regularity 
requirement 
\begin{equation}
  \label{eq:Regularityr0}
  \left(1+ r{\rho_\X'\over\rho_\X}\right) W_\X + r W_\X' - l(l+1) V_\X \sim \O(r^2)\,.
\end{equation}
Regularity of solutions of Poisson's equation (\ref{eq:dPhi}) requires
\begin{equation}
  \label{eq:RegularityPhi}
  \d\Phi \sim \O(r^2)\,\quad\text{for}\quad l> 0\,,
\end{equation}
while for radial oscillations ($l=0$) only $\d\Phi' \sim \O(r)$ is
required. These constraints are automatically satisfied by
(\ref{eq:BCcentre}) for $l\ge2$, but they are stronger requirements
than (\ref{eq:BCcentre}) in the cases $l=1$ and $l=0$. 

\subsubsection{At the surface}

At the outer surface ($r=R$) we need to ensure the continuity of
the gravitational potential $\Phi$, which results (e.g. see
\citet{Ledoux58}) in the boundary condition 
\begin{equation}
  \label{eq:PhiBC}
  \d\Phi'(R) + {l+1\over R}\d\Phi(R) = -\xi^r\, \rho(R-)\,,
\end{equation}
where $\xi^r$ is the radial displacement of the surface, and
$\rho(R-)$ is the \emph{inner} limit of $\rho(r)$,
i.e. $\rho(R-)=\lim_{r\rightarrow R-}\rho(r)$.
In the present work we will only consider stars with vanishing density 
at the surface\footnote{We note that \citet{Lindblom94,Lindblom95}
  have set the right--hand side of (\ref{eq:PhiBC}) to zero in their
  homogeneous model, which is inconsistent if one allows for
  surface displacements $\xi^r$.}  
i.e. $\rho(r)\rightarrow0$.  
The conservation equations (\ref{eq:dmun}) and (\ref{eq:dmuc}) 
contain a (regular) singularity at the surface $r=R$, because 
$\rho'/\rho$ diverges as $\rho\rightarrow0$. Therefore we need an
additional regularity condition that can be obtained by a
Frobenius--type expansion around the surface\footnote{We can
  conveniently use $\mut$ as an effective radial coordinate at the
  surface, because it is well-behaved and monotonic
  (i.e. $\mut(R)=0$ and $\mut'(R)<0$).},i.e. $\mut=0$.  
We first rewrite these equations as
\begin{equation}
  \label{eq:Frob1}
  (r W_\X)' - l(l+1) V_\X = -\left[ r {\rho_\X'\over \rho_\X}W_\X
  + i\freq r^2 {\sum \S_{\X\Y} \d\mut^\Y \over \rho_\X}\right],\;\;
\end{equation}
and using the definition (\ref{eq:DefSXY}) of the matrix $\S_{\X\Y}$
together with the equilibrium condition ${\mut^\n}{}' = \mut^\c{}' =
\mut'$, we can write 
\begin{equation}
  \label{eq:rhoXprime}
  \rho_\X' = k_\X \,\mut'\,,
\end{equation}
where $k_\X$ has been defined in (\ref{eq:DefkX}).
In order to simplify the analysis, we will assume that the equation of
state becomes decoupled in the limit of vanishing density, i.e.
\begin{equation}
  \label{eq:DecoupledEOS}
  P(\mut^\n, \mut^\c) \sim P_\n(\mut^\n)+P_\c(\mut^\c)\,, 
\end{equation}
such that $\rho_\n\sim\rho_\n(\mut^\n)$ and
$\rho_\c\sim\rho_\c(\mut^\c)$, and $k_\X \sim \S_{\X\X}$ in this limit.
In this case we can write the asymptotic form of
(\ref{eq:Frob1}) as   
\begin{equation}
  \label{eq:Frob2}
    (r W_\X)' - l(l+1) V_\X = -{\S_{\X\X}\over\rho_\X}
\left[ r \mut' W_\X + i\freq r^2 \d\mut^\X\right],
\end{equation}
where the expression in brackets is regular for $\mut\rightarrow0$.
The behaviour of $\S_{\X\X}/\rho_\X$ in this limit can be analysed
by writing  $\rho_\X$ in the asymptotic form 
$\rho_\X \sim (\mut^\X)^\alpha$ (with positive $\alpha$), which is
possible due to the assumed decoupling.
Therefore we find $\S_{\X\X}=\partial \rho_\X/\partial\mut^\X\sim
\mut^{\alpha-1}$, and so we have identified the singularity in
(\ref{eq:Frob2}) as 
\begin{equation}
  \label{eq:Frob3}
  {\S_{\X\X}\over\rho_\X} \sim \mut^{-1}\,.
\end{equation}
Regularity of the solution therefore requires the following asymptotic
behaviour at the surface:
\begin{equation}
  \label{eq:Frobenius}
  r \mut' W_\X + i\freq r^2 \d\mut^\X \sim \O(\mut)\,,\quad\text{for}
\quad \X\in\{\n,\,\c\}\,.
\end{equation}

\subsection{Decoupling ``ordinary'' and ``superfluid'' modes?}
\label{sec:DecouplingOrdSf}
In this section we discuss a change of variables that has been used in
several previous studies of oscillations of superfluid neutron stars
\citep{Lindblom94,Lindblom95,Lindblom00,Sedrakian00,andersson01:_dyn_superfl_ns},
namely 
\begin{equation}
  \label{eq:VariablesOrdSf}
  \begin{array}{r l r l}
\d\vDv \equiv & \d\vv_\c - \d\vv_\n \,,\quad & \dbeta \equiv & \d\mut^\c - \d\mut^\n \,,\\
\d\vv \equiv & \xc\,\d\vv_\c + \xn \,\d\vv_\n \,,\quad&
\d\mut \equiv & \xc\,\d\mut^\c + \xn \,\d\mut^\n \,.\\
\end{array}
\label{eq:TraditionalCoordinates}
\end{equation}
This choice of variables is motivated by the intuitive idea that the
additional degrees of freedom of a second fluid should allow for two 
different types of motion, characterised roughly by the two fluids being either
``co-moving'' or ``counter-moving'', and which are sometimes referred
to as ``ordinary'' and ``superfluid'' modes.
By choosing such ``adapted'' coordinates
(\ref{eq:TraditionalCoordinates}), one might hope to
separate, or at least simplify the system of equations, but we will
see that this is generally not the case.

Using the relations (\ref{eq:drhodmu}) and the definition
(\ref{eq:rhoXprime}) of $k_\n$ and $k_\c$, and defining 
$k\equiv k_\n + k_\c$, we can express the density perturbations
$\d\rho_\X$ as 
\begin{equation}
  \label{eq:dRhoX}
  \begin{array}{r l}
    \d\rho_\n = & k_\n \d\mut +\left(\S_{\n\c} - \xc k_\n \right)\dbeta\,,\\
    \d\rho_\c = & k_\c \d\mut +\left(-\S_{\n\c} + \xn k_\c\right)\dbeta\,,\\
  \end{array}
\end{equation}
and therefore
\begin{equation}
  \label{eq:dRho}
  \d\rho = k \d\mut - (\xc k_\n - \xn k_\c)\, \dbeta\,.
\end{equation}
In the case of a spherically symmetric background considered here, we
can use (\ref{eq:rhoXprime}) to express
\begin{equation}
  \xc k_\n - \xn k_\c = - k{\rho\over\rho'} \,\xc'\,,
\end{equation}
and using these variables and relations, we now rewrite the eigenmode
equations  (\ref{equ20})--(\ref{equ24}) in the form
\begin{eqnarray}
k \,\partial_t \d\mut + \vnabla\cdot(\rho\,\d\vv) &=& 
- \xc' \,\left[k{\rho\over\rho'}\,\partial_t \dbeta \right]\,,\label{eq:Ord1}\\
\partial_t \d\vv + \vnabla(\d\mut + \d\Phi) &=&
\;\;\,\xc' \,\left[ \dbeta\,\vec{e}^r\right]\,,\label{eq:Ord2}\\
\nabla^2 \d\Phi -  k \, \d\mut &=& -\xc' \,\left[k{\rho\over\rho'}\,\dbeta\right]\,, 
\label{eq:Ord3}\\
\K\, \partial_t \dbeta + \vnabla\cdot\left( \rho\xn \xc\,\d\vDv\right) &=& 
-\xc'\left[k{\rho\over\rho'}\,\partial_t \d\mut + \rho\,\d v^r
\right]\!,\label{eq:Sf1} \\
\left(1-{2\entr \rho\over\rho_\n \rho_c}\right) \partial_t {\d\!\vDv} +
\vnabla \dbeta &=& \;\;0\,,\label{eq:Sf2}
\end{eqnarray}
where $\vec{e}^r$ is the radial basis vector, and we introduced the
abbreviation $\K \equiv \xc^2 k_\n + \xn^2 k_\c - \S_{\n\c}$.

We see that the ``ordinary''-type of motion ($\d\mut$, $\d\vv$,
$\d\Phi$) does not decouple from the ``superfluid''-type
variables ($\d\vDv$, $\dbeta$) whenever there is \emph{stratification},
i.e. when $\xc'\not=0$ !
This can be understood as follows: while a non--zero relative velocity
$\d\vDv$ can be regarded as a characteristic of a superfluid mode (as
opposed to modes in a single fluid), the chemical equilibrium
deviation $\dbeta$ is generally non--zero even for a single
(but non--barotropic) fluid. 
In a stratified fluid, any general adiabatic motion will drive a
fluid element out of equilibrium, i.e. $\dbeta$ nonzero is not
characteristic for either ``superfluid'' or ``ordinary'' modes
(contrary to claims in \citet{Lindblom94,Lindblom95}), it is a
general feature of modes in stratified fluids, and therefore the
choice of variables (\ref{eq:VariablesOrdSf}) does not lead to a
decoupling of the system in this case. 

However, it is interesting to consider for a moment this special case
of a non--stratified background (which probably never applies in
real neutron stars). Setting the proton fraction $\xc$ to a
constant, we can separate the equations into two decoupled sets. One
system describes  ``ordinary'' modes, namely
\begin{eqnarray}
k \partial_t \d\mut + \vnabla\cdot( \rho\d\vv ) &=& 0\,,\\
\partial_t \d\vv + \vnabla(\d\mut + \d\Phi) &=& 0\,,\\
\nabla^2 \d\Phi -  k \, \d\mut &=&  0\,,
\end{eqnarray}
which are seen to be independent of  the entrainment $\eps$ as well as
of the coupling through the ``bulk'' equation of state,
i.e. $\S_{\n\c}$. The second system of equations governs the
``superfluid'' modes and reads as  
\begin{eqnarray}
\left(k - {\S_{\n\c}\over\xn\xc}\right) \,\partial_t \dbeta + 
\vnabla\cdot(\rho\,\d\vDv) &=& 0\,.\\
\left(1-{2\entr\rho\over\rho_\n\rho_\c}\right)\, \partial_t\, \d\vDv +
\vnabla \dbeta &=& 0\,,
\end{eqnarray}
We see that contrary to the ordinary modes, the superfluid modes do
depend on the coupling through entrainment $\entr$ and the equation of
state, i.e. $\S_{\n\c}$, but they are completely decoupled from the
gravitational perturbation $\d\Phi$, as they leave the total density
unchanged, i.e. $\d\rho=0$.

We have therefore shown that in the non--stratified case there exist
two separate families,  namely ``ordinary'' modes $(\d\vv,\d\mut, \d\Phi, 0, 0)$ and 
``superfluid'' modes $(0,0,0,\d\vDv,\dbeta)$. 
One of our numerical models (see next section) has a constant proton
fraction $\xc$, and we  will see the present analysis confirmed by the
numerical results for this model. In the general stratified case,
however, these two mode families are coupled and such a clearcut
separation is not possible.  

\section{Numerical results}
\label{sec:NumSol}

\subsection{Equation of state: two--constituent polytropes}

We use a simple class of two--constituent equations of state which is
very convenient to explore the properties of a two--fluid system,
namely the following ``generalised polytrope'', defined as\footnote{This
 equation of state has been used previously by
 \citet{comer99:_quasinorm_modes_GR_superfl_NS} and 
\citet{andersson02:_oscil_GR_superfl_NS} to study two--fluid
oscillations in general relativity.} 
\begin{equation}
  \label{eq:2FluidPoly}
  \E^\stat(\rho_\n, \rho_\c) = \kn \,\rho_\n^\Gn + \kc\, \rho_\c^\Gc 
\end{equation}
which simply consists of the sum of two ordinary polytropes.
For regularity of the chemical potentials (\ref{eq:FirstLaw})
in the limit $\rho_\n\rightarrow0$ and $\rho_\c\rightarrow0$, the
polytropic indices must satisfy 
$\Gn\ge1$ and $\Gc \ge1$.
This equation of state allows the explicit inversion 
\begin{equation}
\rho_\n(\mut^\n) = \left( {\mut^\n \over \kn \Gn}\right)^{N_\n}\,,
\;\;\text{and}\;\;
\rho_\c(\mut^\c) = \left( {\mut^\c \over \kc \Gc} \right)^{N_\c}\,,
\label{equDens}
\end{equation}
where we have introduced \mbox{$N_\X \equiv 1/(\GX -1)$}.
We see that in chemical equilibrium, i.e. $\mut^\n=\mut^\c$, 
the two fluids share a common outer surface. The equilibrium
proton fraction $\xc$ can be expressed as  
\begin{equation}
  \label{eq:ProtonFraction}
  \xc = \left( 1 + { (\Gc \kc)^{N_\c} \over (\Gn \kn)^{N_\n} }
    \, \mut^{N_\n - N_\c} \right)^{-1} \,,
\end{equation}
which shows that the proton fraction is constant whenever $\Gn=\Gc$,
while the behaviour in the case of different indices falls
into the two categories:
\begin{eqnarray}
0 \stackrel{\mut\rightarrow\infty}{\longleftarrow} \xc 
\stackrel{\mut \rightarrow 0}{\longrightarrow} 1\,,\quad\text{for}\quad 
\Gn < \Gc \,,\\
1 \stackrel{\mut\rightarrow\infty}{\longleftarrow} \xc 
\stackrel{\mut \rightarrow 0}{\longrightarrow} 0\,,\quad\text{for}\quad 
\Gn > \Gc \,.
\end{eqnarray}

\subsection{Calculating the background models}
\label{sec:NumBgSolution}

The equilibrium background solution is determined by equation
(\ref{eq:BgConf}), together with the regularity requirement
$\mut'(0)=0$, and the boundary condition of vanishing pressure at the
surface, i.e.\footnote{Strictly 
speaking $\mu$ is determined only up to a constant, which is usually
fixed such that $\mut$ vanishes together with the pressure.}
$\mut(1)=0$, where the surface of the static background star is
situated at $R=1$ in the natural units defined in Sect.~\ref{sec:StaticBg}.  
Using the equation of state we can express $\rho = \rho(\mut)$ in
chemical equilibrium, and therefore equation (\ref{eq:BgConf}) can be
written as the following nonlinear eigenvalue problem, 
\begin{equation}
  \label{eq:BgEigenvaluePb}
  \mut'' + {2\over r} \mut = \lambda \,\rho(\mut)\,,
\end{equation}
where the (dimensionless) eigenvalue $\lambda$ is given by
\begin{equation}
  \label{eq:BackgroundEigenvalue}
  \lambda = 4\pi G {\rho_0 \over \mut(\rho_0)}  R^2 \,.
\end{equation}
The eigenvalue $\lambda$ represents the a--priori unknown
radius of the star, and thereby the actual values of our
natural units for a given central density $\rho_0$. 
The method used here to solve this equation is to iterate a linear
eigenvalue problem that converges to the solution of
(\ref{eq:BgEigenvaluePb}). This can be done by solving in step
$k+1$ the following linear eigenvalue problem,
\begin{equation}
  \label{eq:BgLinEigenvaluePb}
  \mut''_{k+1} + {2\over r}\mut'_{k+1} = -\lambda_{k+1} \left(
  {\rho(\mut_k)\over \mut_k}\right)\, \mut_{k+1}\,,
\end{equation}
where $\mut_k$ is the solution of the of the previous step $k$.
The equation in each step is solved using the spectral linear
eigenvalue solver  package LSB developed by L.~Valdettaro and M.~Rieutord.
With a resolution of 40 Chebychev polynomials and a Gauss--Lobatto
collocation method, this iteration converges to about machine
precision (i.e. $\sim10^{-15}$ relative difference between successive
steps) in about $20$ steps. Another practical 
advantage of this method is that we calculate the background
quantities on the same Gauss--Lobatto grid we use for the
numerical code for the eigenmodes. Even when using a different
resolution for the eigenmode--calculation, the Chebychev expansion
provides a canonical $\C^\infty$ interpolation which allows us to
easily ``re-grid'' the background solution. 
\begin{figure*}
  \centering
  \includegraphics[width=17cm,clip]{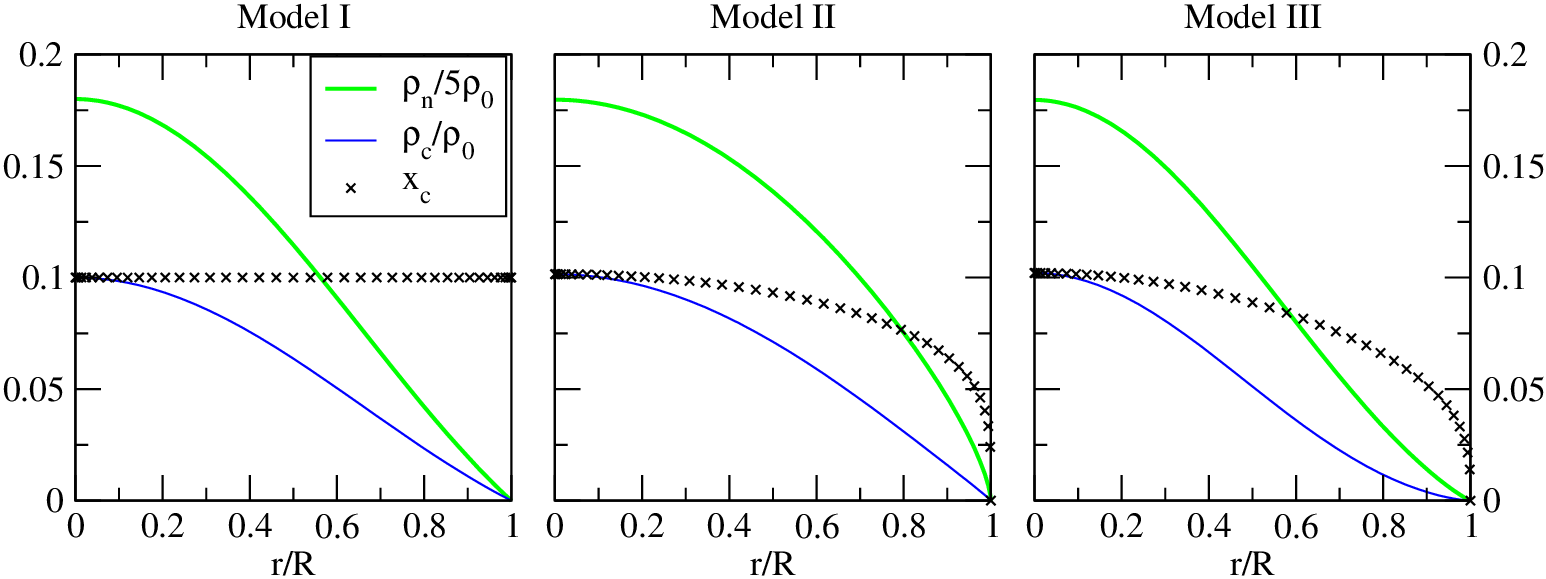}
  \caption{Density profiles and proton fraction $\xc$ of the
  background models I, II and III as defined in table
  \ref{tab:BackgroundModels}. $\rho_0$ is the central density. In this
  figure we have divided the neutron density by 5 in order to obtain
  similar magnitudes for the different curves.}
  \label{fig:BackgroundModels}
\end{figure*}

In principle we can solve the background for any given equation of
state, but we restrict our attention in this work to the class of
two--constituent polytropes (\ref{eq:2FluidPoly}).The reason for this
choice is their analytic simplicity and because our main emphasis is
to clarify the qualitative properties of superfluid neutron stars
rather than to construct a physically very precise model. This would
in any case be quite impossible in a Newtonian description because of
the neglect of relativistic effects, in addition to the important
uncertainties in our current knowledge of the equation of state of
neutron stars.  

In the numerical analysis we consider the three different background 
models defined in table~\ref{tab:BackgroundModels}, and which are
represented  in Fig.~\ref{fig:BackgroundModels}. 
These three models 
correspond to three different types of behaviour at the surface.
In the case of Model~I one can easily find the background solution
analytically, as shown in \citet{prix02:_slowl_newton},
which allows us to check the numerical method of calculating the
background, and we find a maximal relative error of $10^{-14}$ between
the numerical and the analytic solution for model~I. 
Model~II represents a generic ``stiff'' model similar to those used in
\citet{comer99:_quasinorm_modes_GR_superfl_NS} and \citet{andersson02:_oscil_GR_superfl_NS}, which has infinite density
gradients at the surface. Model~III  is of a ``soft'' type with
vanishing density gradients at the surface. These different types of
behaviour at the surface are quite analogous to the case of the usual
one--constituent polytropes for different polytropic indices.  
\begin{table}
  \centering
  \begin{tabular}{l| l l l l| l l l l }
    \rule[-0.3cm]{0.0cm}{0.8cm} & $\Gn$ & $\Gc$ & $\kn^a$  & $\kc^a$ & ${\rho_0\over\rho_\nuc}$ & $\xc(0)$ & ${M\over\Msol}$ & ${R\over\km}$ \\\hline
    \rule[-0.3cm]{0.0cm}{0.8cm} 
I   & 2.0 & 2.0 & 0.01 &  0.09 &  10.83 & 0.10 & 1.414 & 10.7 \\ 
II  & 2.5 & 2.1 & 0.01 &  0.20 &  3.20 & 0.10 & 1.440 & 14.3 \\ 
III & 1.9 & 1.7 & 0.01 &  0.09 &  18.65 & 0.10 & 1.412 & 9.4  \\
  \end{tabular}
  \caption{Parameters describing the  background models I, II and III, 
    based on the two--constituent polytrope (\ref{eq:2FluidPoly}).
    Further shown are the central mass density $\rho_0$, the
    central proton fraction $\xc(0)$, the total mass $M$ and the
    radius~$R$. 
    $^a$~The~units~of~the~coefficients~$\k_\X$~are~$\c^2{\rho_\nuc}^{1-\GX}$,~where~$\rho_\nuc=1.66\times10^{14}$~g~cm$^-3$.}
  \label{tab:BackgroundModels}
\end{table}
We note that model~I is the only non--stratified model, because it has
a constant proton fraction $\xc$ (and finite density gradients at  the
surface), while models II and III have a non--zero composition
gradient, i.e. $\xc'\not=0$, as expected from (\ref{eq:ProtonFraction}).  

For easier comparison of the frequencies given in the next section 
in units of $\sqrt{4\pi G \rho_0}$, we provide in
table~\ref{tab:UnitConversions} the conversion factors into three
important systems of units, namely the SI unit Hz, the ``Cox'' units
$\sqrt{G M/R^3}$ (variants thereof, like those used by
\citet{Lindblom94}, only differ by a constant factor) and the
``geometric'' units $c^3/G M$ typically used in general relativity   
\citep{comer99:_quasinorm_modes_GR_superfl_NS,andersson02:_oscil_GR_superfl_NS}. 
\begin{table}
  \centering
  \begin{tabular}{c || c| c| c}
            & I     & II   &  III \\\hline
  kHz               &  38.8318047 & 21.1116405 &  50.9444446  \\
    $\sqrt{GM/R^3}$ &  3.14159265 & 2.61829313 &  3.40801371  \\
    $c^3/GM$        &  0.270521502 & 0.149723677 & 0.354013053 \\
  \end{tabular}
  \caption{Conversion factors for our frequency units ($\sqrt{4\pi
      G\rho_0}$) into different systems of units (SI,
      \citet{cox76:_nonrad} and ``geometrised'' units) for models I, II and
      III.}  
    \label{tab:UnitConversions}
\end{table}

\subsection{The two--fluid oscillation modes}
\label{sec:NumSfEigenmodes}

The eigenmode equations (\ref{eq:Wn})-(\ref{eq:dPhi}) together with
the boundary conditions of Sect.~\ref{sec:BoundaryConditions} form a
linear eigenvalue system which we solve numerically using the
spectral solver of the LSB--package. 
The convergence of the results was determined by increasing the
resolution starting from $40$ Chebychev polynomials up to $80$, and we
found the changes in frequency decrease very quickly to about 
$10^{-9}$ (or better), which is why we give the frequencies with nine
decimals in tables \ref{tab:Eigenvalues0} to \ref{tab:Eigenvalues2},
corresponding roughly to the convergence achieved by the numerical
method.  This is for future reference and comparison, not because
these frequencies represent a physically measurable prediction in any
sense.

\subsubsection{Eigenmodes of ``locally uncoupled'' fluids}

In this section we consider the case of zero entrainment,
i.e. $\entr=0$ and $\eps_\n=\eps_\c=0$. 
We refer to this situation as ``locally uncoupled'' fluids, as it is
important to note that the two fluids are nevertheless coupled 
``globally'' through the perturbation of the gravitational potential
$\d\Phi$ and (\ref{eq:dPhi}). 
We consider the cases of radial ($l=0$), dipolar ($l=1$) and
quadrupolar ($l=2$) oscillations, which differ qualitatively in some
properties and boundary conditions (see
Sect.~\ref{sec:BoundaryConditions}), while all higher $l$ cases are 
qualitatively very similar to $l=2$.
The lowest eigen-frequencies for these three values of $l$ are shown
in tables~\ref{tab:Eigenvalues0}, \ref{tab:Eigenvalues1} and
\ref{tab:Eigenvalues2} respectively.
We label these modes in analogy to the one--fluid case as f- and p-
modes, and group them in pairs where the lower frequency mode is labelled
as ``$\ord$'' and the higher frequency one as ``$\sf$''. 
The pairs of p--modes are indexed in the order of increasing
frequency.  We emphasise that this labelling is a pure convention,
as one can generally \emph{not} say that these modes would be either
co-- or counter-moving, or that the subscript would exactly reflect
the number of radial nodes.
\begin{figure*}
  \centering
  \includegraphics[width=17cm,clip]{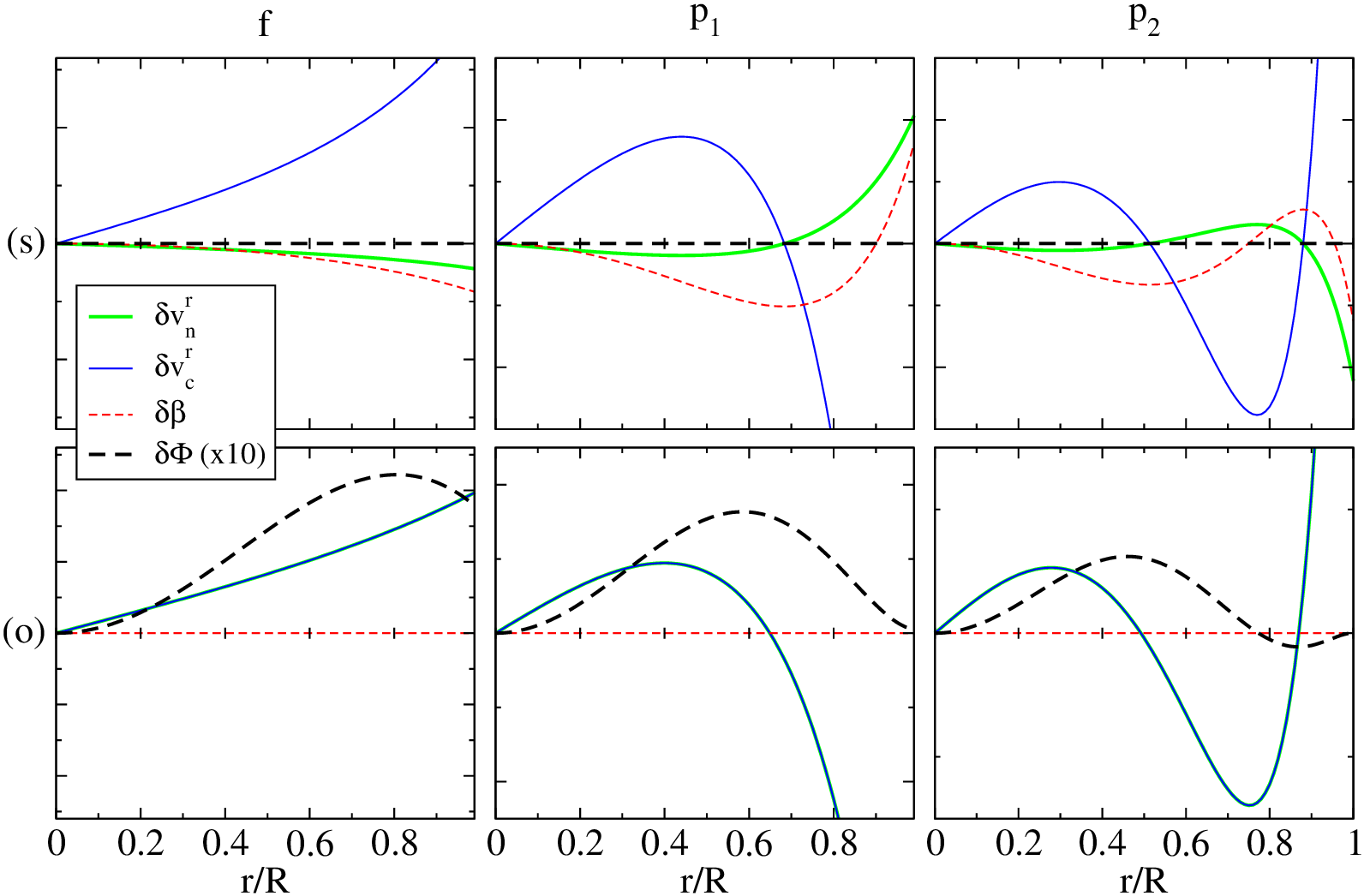}
  \caption{The first three quadrupolar ($l=2$) eigenmode doublets
  ($\ord$)/($\sf$) for the background model I. In the case of the 
    $\ord$ -- modes, the two fluids are comoving to numerical
    precision, therefore the $\d v_\n$ and $\d v_\c$ curves can not be
    seen separately.} 
\label{fig:EigenmodesModelI}
\end{figure*}

Let us first consider the special separable case of the
non--stratified model~I. 
The first three pairs of eigenfunctions are presented in
Fig.~\ref{fig:EigenmodesModelI}, and we see that in the ``$\ord$'' modes
the two fluids are comoving, resulting in a non--zero
$\d\Phi$, and they also remain in strict chemical equilibrium,
i.e. $\dbeta=0$. These ``ordinary'' modes are actually identical to
the normal--fluid modes of the same background (see
Sect.~\ref{sec:Num1fEigenmodes}). 
In the case of the $\sf$--type modes the two fluids are
counter--moving in exactly such a way that the total density remains
constant, i.e. $\d\rho=0$ and therefore $\d\Phi=0$, while the fluids
are driven out of chemical equilibrium, i.e. $\dbeta\not=0$. The
number of radial nodes in $\d\vv^r$ is the same for the $\ord$ and
$\sf$ modes, and corresponds exactly to their index. 
All these results confirm the analytic predictions for non--stratified
models in Sect.~\ref{sec:DecouplingOrdSf}. 
\begin{figure*}
  \centering
  \includegraphics[width=17cm,clip]{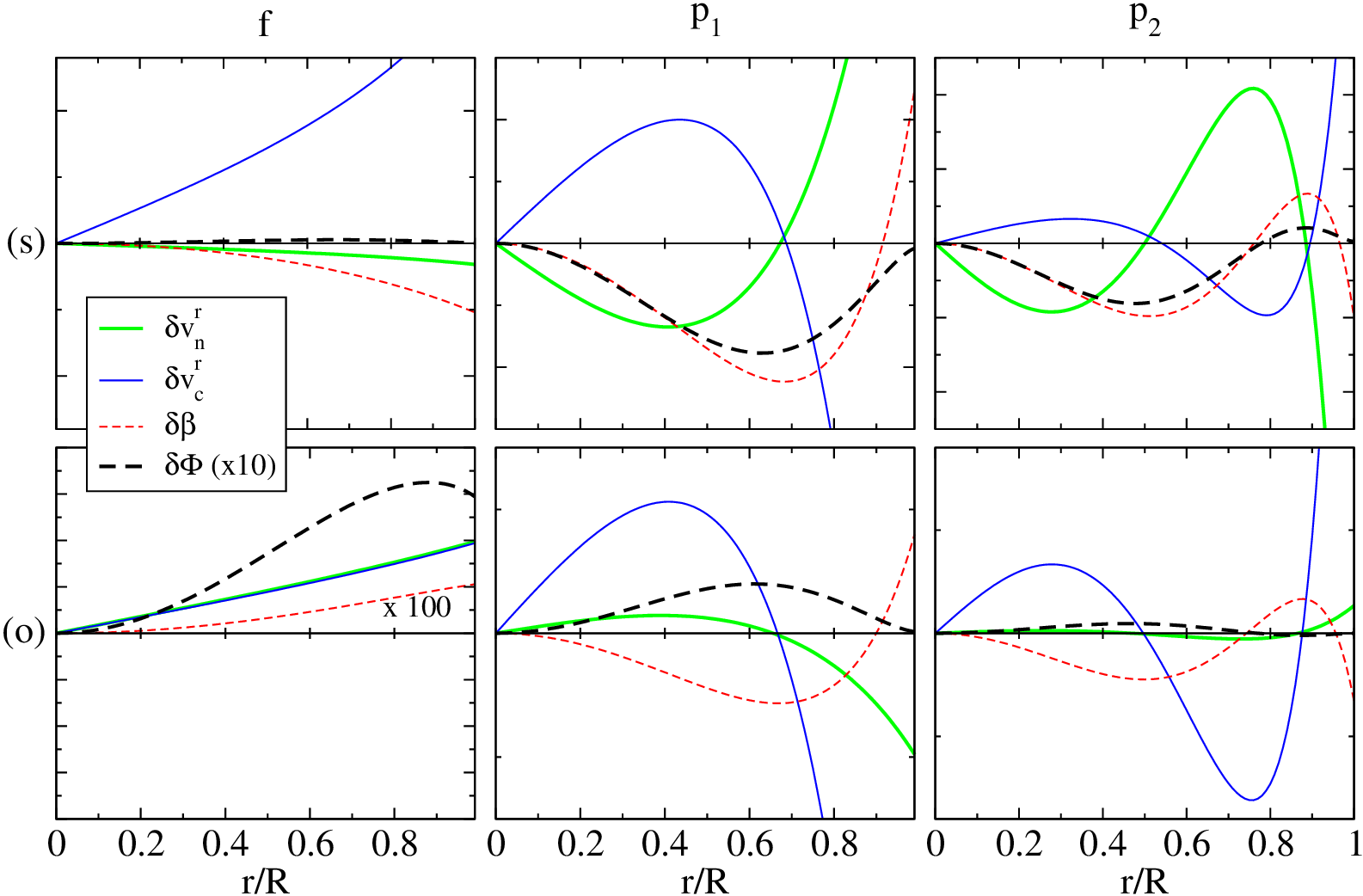}
  \caption{The first three quadrupolar ($l=2$) eigenmode doublets
  ($\ord$)/($\sf$) for the background model II. The value of $\dbeta$
  in the lower left figure has been multiplied by 100, in order to
  show that it is non--zero.}
\label{fig:EigenmodesModelII}
\end{figure*}
\begin{figure}
  \centering
  \includegraphics[width=\hsize,clip]{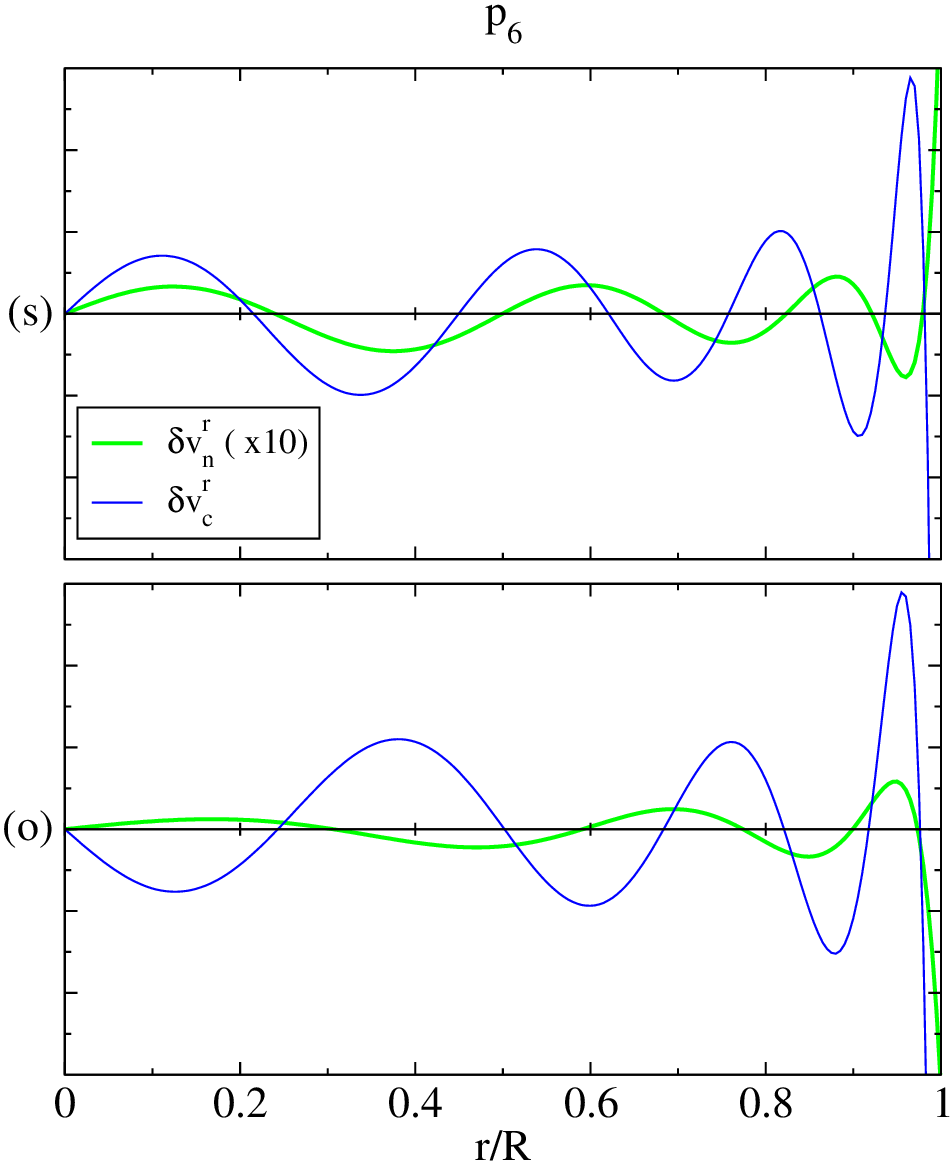}
  \caption{The p$_6$ eigenmode doublet ($\ord$)/($\sf$) (for $l=2$)
  for the background model II. The radial velocity $\d v^r_\n$ has
  been multiplied by 10 for better visibility. Neither mode is co-- or
  counter-moving, and the number of radial nodes differs between the
  two fluids.}
\label{fig:p6Eigenmode}
\end{figure}
However, it would be wrong to assume that these
properties are generally true for superfluid oscillations.
Stratification makes this picture more complex, even in the case of
locally uncoupled fluids considered in this section. 
If we look at the first three pairs of eigenfunctions for the model~II
in Fig.~\ref{fig:EigenmodesModelII}, we see that 
the ``$\ord$'' modes are not comoving at all (only the f$^\ord$ is
nearly comoving), and they have non--zero $\dbeta$,
while the ``$\sf$'' modes have non--zero $\d\rho$ and $\d\Phi$.
One can not say either that the relative amplitude of $\dbeta$ would
be different between the $\ord$-- and $\sf$-- cases, as
\citet{Lindblom94} wrongly induced from the properties of the
f$^\ord$--mode, the only eigenmode they presented.

In the case of the low--order modes presented in
Fig.~\ref{fig:EigenmodesModelI}, the number of radial nodes still
seems to correspond to the index, and the fluids are roughly in
opposite phase in the $\sf$-- modes, while they are approximately in
phase for the $\ord$--modes. Even this, however, is not true in
general, as can be seen in Fig.~\ref{fig:p6Eigenmode} which shows the
higher order p$_6$--modes. 
In this case neither $\ord$-- nor $\sf$-- are dominantly in phase or
in opposite phase, and the two radial velocities have different numbers of
radial nodes. Therefore neither the index nor the $\ord$/$\sf$ label
bear any reliable information about the properties of the modes. 
This behaviour is possible because the that this eigenvalue
problem is not of Sturm--Liouville type except in the non--stratified case.
\begin{table}
  \centering
  \begin{tabular}{c || c | c |  c}
    $l=0$ &   $\freq_I$  & $\freq_{II}$ & $\freq_{III}$ \\ \hline\hline
    f$^\ord$   &  0.616\,801\,012 &  0.860\,501\,159 &  0.539\,820\,916\\
    f$^\sf$    &  0.825\,395\,141 &  1.004\,218\,360 &  0.713\,202\,951\\\hline
    p$_1^\ord$ &  1.272\,153\,763 &  1.650\,440\,676 &  1.114\,032\,342\\
    p$_1^\sf$  &  1.398\,557\,067 &  1.780\,808\,966 &  1.186\,350\,906\\\hline
    p$_2^\ord$ &  1.855\,852\,617 &  2.326\,264\,710 &  1.582\,750\,279\\
    p$_2^\sf$  &  1.949\,822\,942 &  2.573\,698\,536 &  1.675\,075\,045\\\hline
    p$_3^\ord$ &  2.418\,457\,671 &  2.985\,429\,110 &  2.018\,290\,454\\
    p$_3^\sf$  &  2.493\,326\,179 &  3.352\,297\,316 &  2.169\,556\,932\\\hline
    p$_4^\ord$ &  2.970\,977\,248 &  3.638\,960\,946 &  2.445\,611\,455\\
    p$_4^\sf$  &  3.033\,169\,591 &  4.122\,179\,788 &  2.658\,934\,809\\\hline
  \end{tabular}
  \caption{Frequency spectrum of radial eigenmodes ($l=0$) for models
    I, II and III in natural units $\freq_0=\sqrt{4\pi G
    \rho_0}$.}
\label{tab:Eigenvalues0}
\end{table}
\begin{table}
  \centering
  \begin{tabular}{c || c | c |  c}
    $l=1$ &   $\freq_I$  & $\freq_{II}$ & $\freq_{III}$ \\ \hline\hline
    f$^\ord$   &   $0^a$ & $0^a$ & $0^a$ \\
    f$^\sf$    &   0.389\,835\,134 &  0.440\,176\,989 &  0.387\,577\,390\\\hline
    p$_1^\ord$ &   0.898\,011\,966 &  1.191\,707\,859 &  0.798\,353\,611\\
    p$_1^\sf$  &   1.040\,570\,747 &  1.280\,907\,920 &  0.895\,985\,737\\\hline
    p$_2^\ord$ &   1.518\,621\,841 &  1.930\,140\,941 &  1.317\,638\,486\\
    p$_2^\sf$  &   1.621\,023\,028 &  2.105\,369\,668 &  1.384\,086\,979\\\hline
    p$_3^\ord$ &   2.099\,218\,641 &  2.609\,579\,518 &  1.770\,672\,472\\
    p$_3^\sf$  &   2.179\,811\,073 &  2.908\,118\,483 &  1.887\,282\,252\\\hline
    p$_4^\ord$ &   2.662\,623\,840 &  3.274\,392\,031 &  2.206\,080\,421\\
    p$_4^\sf$  &   2.729\,000\,908 &  3.692\,629\,614 &  2.385\,023\,708\\\hline
  \end{tabular}
  \caption{Frequency spectrum of dipolar eigenmodes ($l=1$) for models
I, II and III in natural units $\freq_0=\sqrt{4\pi G \rho_0}$.
$^a$~The~zero~frequency~modes~correspond~to~the~analytic result of a
constant displacement field $\vec{\xi}$. This solution can not be
produced by our code because the equations have been expressed in
terms of velocities instead of displacements.} 
\label{tab:Eigenvalues1}
\end{table}
\begin{table}
  \centering
  \begin{tabular}{c || c | c |  c}
    $l=2$ &   $\freq_I$  & $\freq_{II}$ & $\freq_{III}$ \\ \hline\hline
    f$^\ord$   & 0.390\,550\,961 &  0.424\,294\,338 & 0.376\,662\,787\\
    f$^\sf$    & 0.526\,990\,499 &  0.604\,904\,572 & 0.516\,636\,947\\\hline
    p$_1^\ord$ & 1.101\,827\,434 &  1.423\,800\,575 & 0.980\,798\,266\\
    p$_1^\sf$  & 1.206\,881\,695 &  1.514\,568\,856 & 1.039\,415\,772\\\hline
    p$_2^\ord$ & 1.723\,444\,868 &  2.164\,156\,378 & 1.478\,705\,720\\
    p$_2^\sf$  & 1.806\,873\,473 &  2.380\,237\,199 & 1.557\,015\,636\\\hline
    p$_3^\ord$ & 2.310\,315\,782 &  2.858\,007\,106 & 1.932\,213\,886\\
    p$_3^\sf$  & 2.379\,342\,400 &  3.200\,741\,221 & 2.072\,313\,682\\\hline
    p$_4^\ord$ & 2.879\,785\,468 &  3.533\,905\,837 & 2.372\,551\,143\\
    p$_4^\sf$  & 2.938\,448\,588 &  3.997\,594\,591 & 2.576\,350\,147\\\hline
  \end{tabular}
  \caption{Frequency spectrum of quadrupolar eigenmodes ($l=2$) for models
    I, II and III in natural units $\freq_0=\sqrt{4\pi G \rho_0}$.}
  \label{tab:Eigenvalues2}
\end{table}
An interesting fact to notice in tables~\ref{tab:Eigenvalues0}
to \ref{tab:Eigenvalues2} is that the fundamental modes (f$^\ord$
and f$^\sf$) are the lowest frequency modes in the spectrum,
in other words there are no g--modes present (which usually lie
far below the f--mode) in these superfluid models. This confirms
the numerical findings by \citet{Lee95} and the local analysis of
\citet{andersson01:_dyn_superfl_ns}.

The absence of g-modes can be made clearer when acoustic modes and
surface gravity modes are filtered out. The latter modes are easily
removed by suppressing surface motions and imposing therefore $W_X=0$ at
the star surface. Acoustic modes, on the other hand, are filtered out by
using the so-called anelastic approximation which makes an expansion in
powers of the Brunt-V\"ais\"al\"a frequency \cite[see][]{DR01,RD02}.
Using this approximation mass conservation now reads
\begin{equation}
\vnabla\cdot(\rho_n\d\vv_\n) = 0\quad {\rm and}\quad
\vnabla\cdot(\rho_c\d\vv_\c) = 0
\label{massc}
\end{equation}
instead of \eqq{eq:EigenvalueSystem2Da} and \eqq{mass2}. In the case we have considered,
i.e. that of no entrainment ($\alpha=0$), the equations of motions read:
\begin{eqnarray}
  \vnabla( \d\mut^\n + \d\Phi ) &=& - i\freq \d\vv_\n \,,\\
  \vnabla( \d\mut^\c + \d\Phi ) &=& - i\freq \d\vv_\c \,,\\
  \nabla^2 \d\Phi &=&  k_\n \, \d\mut^\n  + k_\c\, \d\mut^\c\,,
\end{eqnarray}
Using \eqq{massc}, we eliminate the velocities and are left with the system:
\begin{eqnarray}
  \vnabla\cdot\lp\rho_n\vnabla(\d\mut^\n + \d\Phi )\rp &=& 0\,,\\
  \vnabla\cdot\lp\rho_c\vnabla(\d\mut^\c + \d\Phi )\rp &=& 0\,,\\
  \nabla^2 \d\Phi &=&  k_\n \, \d\mut^\n  + k_\c\, \d\mut^\c\,,
\end{eqnarray}
where we see that the mode frequency has disappeared. Using the boundary
conditions, it turns out that the only solution is
$\d\mut^\c=\d\mut^\n=\d\Phi=0$, thus showing
that no eigenmode exists when acoustic and surface gravity modes are
filtered out.

However, the presence of g--modes due to chemical composition gradients
in normal--fluid neutron star models has been pointed out by
\citet{reisenegger92}, and their possibly observable excitation
in a coalescing binary neutron star has been discussed by
\citet{reisenegger94:_excit} and \citet{lai94:_reson}.  We will see in
Sect.~\ref{sec:Num1fEigenmodes} that these predicted composition
g--modes do indeed appear in the normal--fluid case.  In principle the
presence or absence of these modes could therefore be used as a possibly
observable indicator for superfluidity in neutron stars.

\subsubsection{The effect of coupling by entrainment}

In this section we study the dependence of the mode-frequencies
and properties on the coupling by entrainment. Obviously, we only
need to specify \emph{one} entrainment function, $\eps_\c$ say, as
$\eps_\n$ is then determined by (\ref{eq:EpsDef}). 
Because the uncertainties and differences of the ``realistic'' models
for $\eps_\c$ provided by nuclear physics calculations so far are
still considerable, we chose the simplest entrainment model, namely
$\eps_\c=\eps$ being a constant. The value of this constant $\eps$ can
be related to the proton effective mass $m_\p^*$
(\ref{eq:EffectiveMassEntr}), and is roughly constrained from the 
nuclear physics calculations
\citep{chao72:_proton_superfl,sjoberg76:_effect_mass,baldo92:_superfl_neutr_star_matter,borumand96:_superfl_neutr_star_matter} 
to lie in the range $0.3\leq\eps\leq 0.7$. We nevertheless consider
the broader range between $-0.8\leq\eps\leq 0.8$ to demonstrate the
qualitative behaviour more clearly. This will also show that the
``locally uncoupled'' case $\eps=0$ (considered in the previous section)
is not special, contrary to what one might have expected. 
\begin{figure*}
  \centering
  \includegraphics[width=17cm,clip]{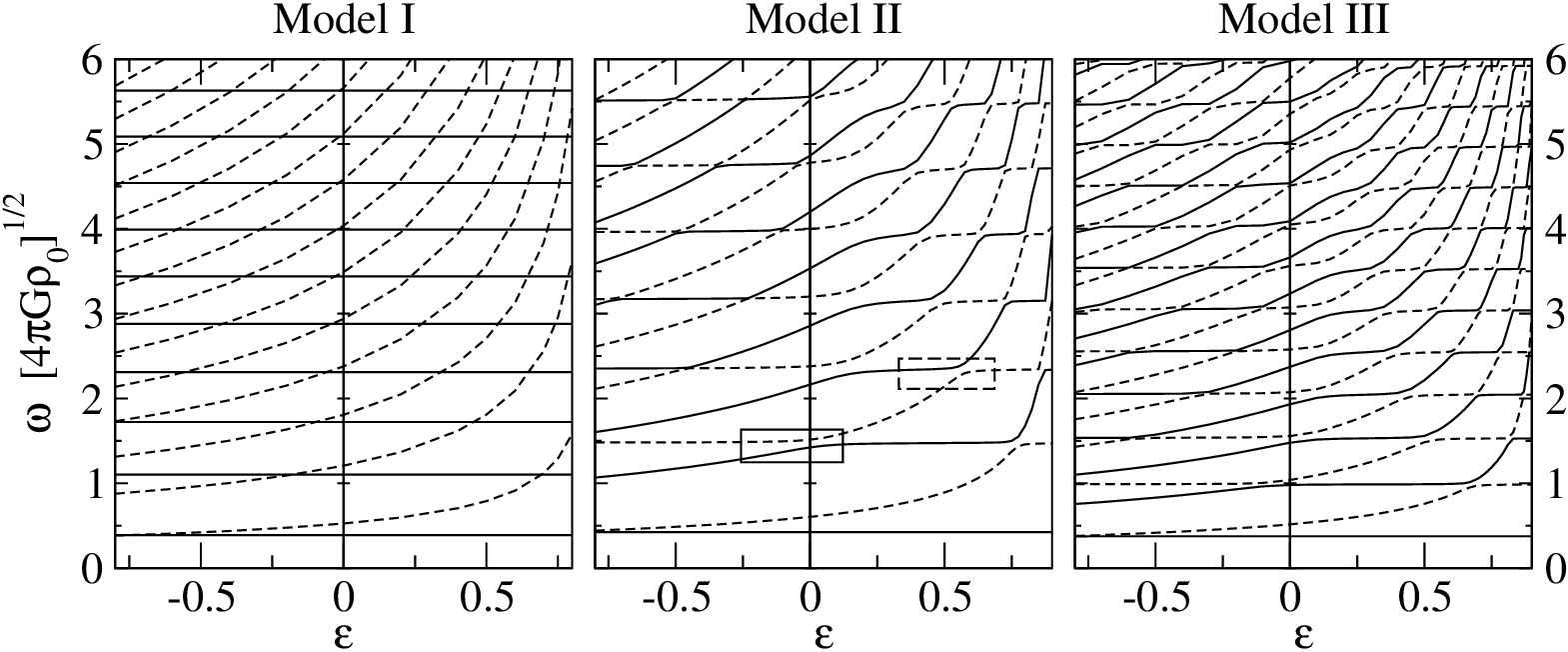}
  \caption{Eigenmode frequencies $\freq$ (for $l=2$) as functions of
    entrainment $\eps$ for models I, II and III. The full and dashed
    lines represent the frequencies of the modes denoted respectively
    by $\ord$ and $\sf$ in table \ref{tab:Eigenvalues2}.
    The avoided crossings marked by the full and the dashed box are
    represented in terms of the eigenfunctions in
    Figs.~\ref{fig:AvoidedModeCrossing11} and
    \ref{fig:AvoidedModeCrossing12}. We note that for
    models~II and III all crossings shown are avoided.} 
  \label{fig:Crossings}
\end{figure*}

\begin{figure}
  \centering
  \includegraphics[width=\hsize,clip]{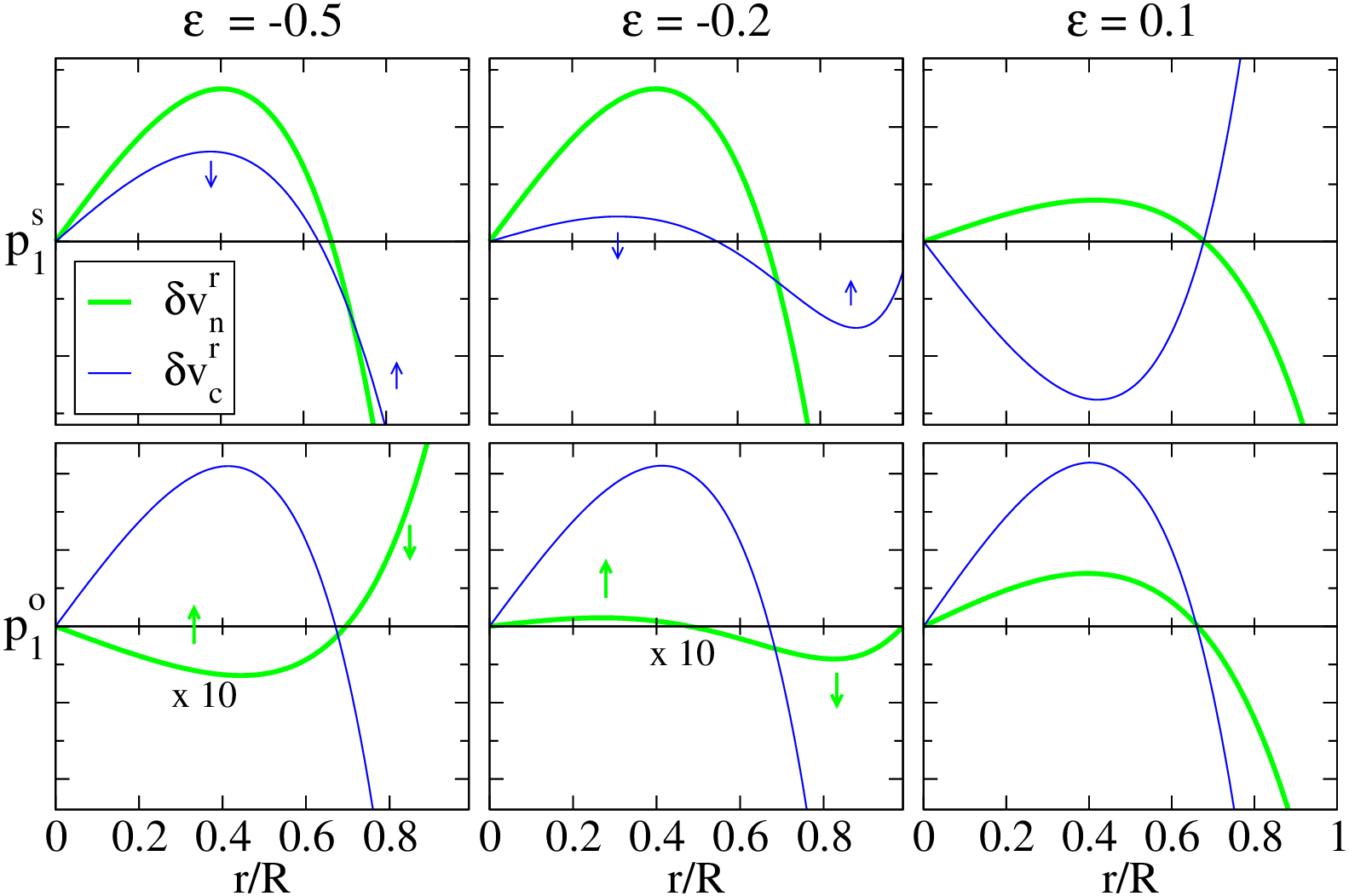}
  \caption{Avoided crossing between the p$_1^\ord$ and the p$_1^\sf$
    modes of Model II (for $l=2$), indicated by the box in
    Fig.~\ref{fig:Crossings}. The three columns show the corresponding
    eigenfunctions for three different values of entrainment. The
    small arrows indicate the ``evolution'' of the eigenmode when
    increasing $\eps$. The $\d v_\n^r$ curves in the two lower left
    figures have been magnified by a factor of $10$.} 
  \label{fig:AvoidedModeCrossing11}
\end{figure}

\begin{figure}
  \centering
  \includegraphics[width=\hsize,clip]{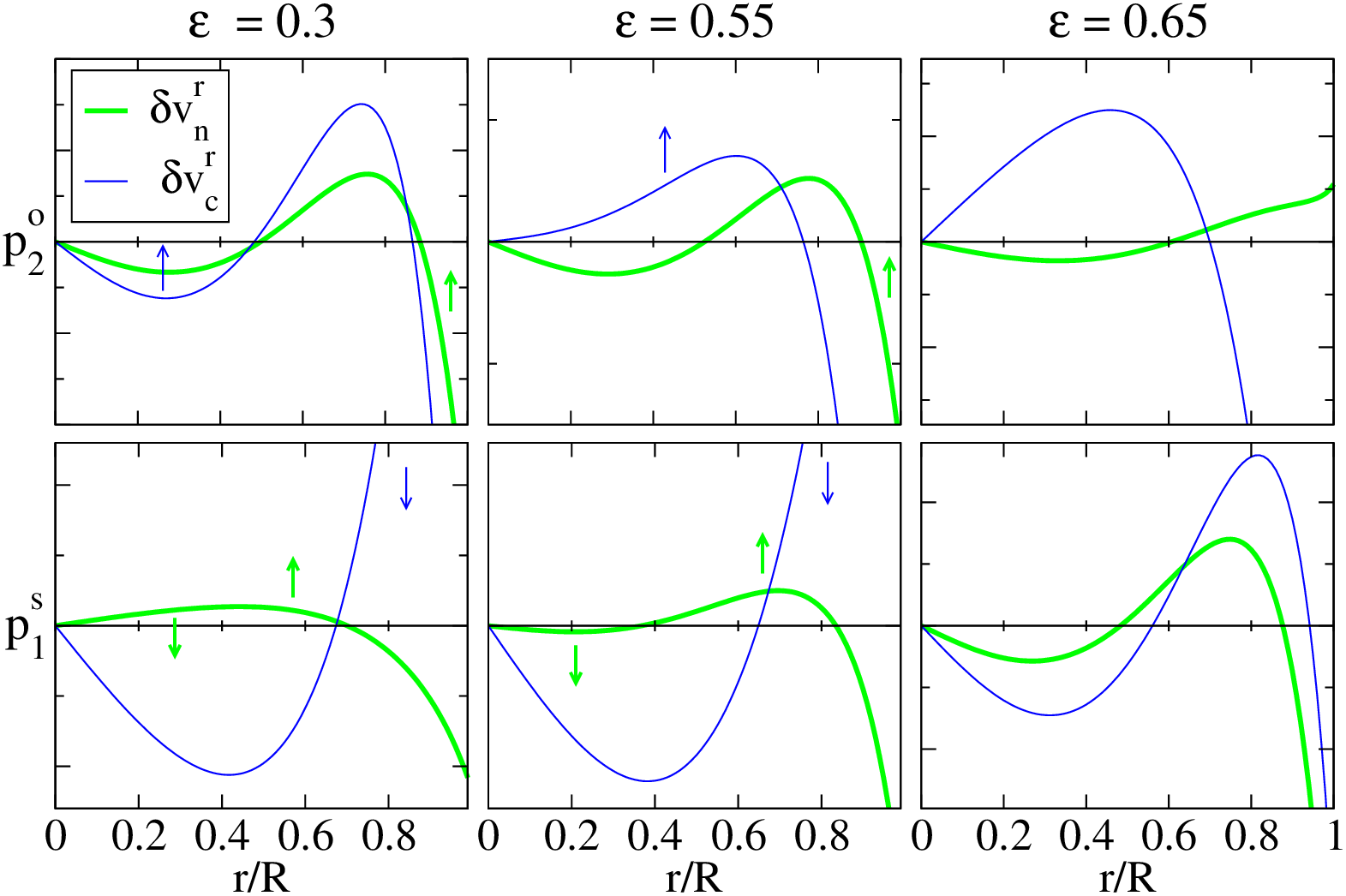}
  \caption{Avoided crossing between the p$_1^\sf$ and the p$_2^\ord$
    modes of Model II ($l=2$), indicated by the hatched box in
    Fig.~\ref{fig:Crossings}. The three columns show the corresponding
    eigenfunctions for three different values of entrainment. 
    The small arrows indicate the ``evolution'' of the eigenmode when
    increasing $\eps$.} 
  \label{fig:AvoidedModeCrossing12}
\end{figure}
The results for the mode-frequencies as functions of $\eps$ for the
three background models are represented in Fig.~\ref{fig:Crossings}.
In the case of the non--stratified model~I, we observe the predicted
(sect.~\ref{sec:DecouplingOrdSf}) decoupling, and in particular the
independence of the ``ordinary''--type modes of entrainment. Because
of this decoupling the respective frequencies of the two mode families
can simply cross each other when $\eps$ is varied. 
In the generic stratified models (model II and III), the
modes of the doublets are coupled and avoided crossings result
when mode--frequencies come to close to each other, as also found
recently by \citet{andersson02:_oscil_GR_superfl_NS}.
In this process of avoided crossing the two modes seem to exchange
some of their respective properties of being dominantly ``co-'' 
or ``counter-moving'', as can be seen in
Fig.~\ref{fig:AvoidedModeCrossing11}), and they also can exchange
their number of radial nodes, as we see in the avoided crossing of the
p$_1^\sf$ and p$_2^\ord$ in Fig.~\ref{fig:AvoidedModeCrossing12}. 

Another important conclusion can be drawn from
Fig.~\ref{fig:Crossings}, namely that the ``locally uncoupled'' case
$\eps=0$ discussed in the previous section does not represent a
special case in any respect, because the two fluids are \emph{always}
coupled through $\d\Phi$. The effect of $\eps$ is simply to
change the coupling, but no configuration is completely
uncoupled. On can see in Fig.~\ref{fig:Crossings} that several avoided
crossings happen practically at $\eps=0$, which is the case in
particular for the p$_6$-modes of model II presented in
Fig.~\ref{fig:p6Eigenmode}.

\subsection{The one--fluid case: recovering the g--modes}
\label{sec:Num1fEigenmodes}

Following the discussion in sect.~\ref{sec:SingleFluidOsc}, the
one--fluid case is defined by $\d\vv_\n = \d\vv_\c \equiv \d\vv$.  
We only have one Euler equation in this case, which in the
harmonic decomposition (\ref{eq:HarmonicBasis}) has the two components 
\begin{eqnarray}
  \xn\, {\d\mut^\n}' + \xc \,{\d\mut^\c}' + \d\Phi' &=& -{i\freq\over r}\, W\,,\\
  \xn\, \d\mut^\n\, + \xc \,\d\mut^\c\, + \d\Phi\,  &=& -i\freq \,V\,.
\end{eqnarray}
These two equations replace (\ref{eq:Wn})--(\ref{eq:Vc}), while the
remaining equations of this system are unchanged (subject to the
substitutions $W_\n=W_\c=W$ and $V_\n=V_\c=V$). The eigen-frequencies
of this system are shown in table~\ref{tab:1FluidEigenvalues}, where we
see the presence of composition g--modes as expected for all
models with stratification. This is consistent with the prediction by
\citet{reisenegger92} and the numerical findings of \citet{Lee95}.
\begin{table}
  \centering
  \begin{tabular}{c || c | c |  c}
    $l=2$ &   $\freq_I$  & $\freq_{II}$ & $\freq_{III}$ \\ \hline\hline
   \dots  &         --       &     \dots & \dots \\
    g$_4$ &         --       &  0.012\,105\,268 &  0.011\,489\,709 \\
    g$_3$ &         --       &  0.015\,003\,335 &  0.014\,157\,505 \\
    g$_2$ &         --       &  0.019\,814\,997 &  0.018\,492\,575 \\
    g$_1$ &         --       &  0.029\,631\,110 &  0.026\,880\,058 \\\hline
    $f$   &  0.390\,550\,961 &  0.424\,310\,492 &  0.376\,717\,911 \\\hline
    p$_1$ &  1.101\,827\,434 &  1.477\,988\,230 &  0.988\,324\,062 \\
    p$_2$ &  1.723\,444\,868 &  2.348\,478\,094 &  1.533\,337\,250 \\
    p$_3$ &  2.310\,315\,782 &  3.163\,853\,143 &  2.050\,348\,013 \\
    p$_4$ &  2.879\,785\,468 &  3.954\,289\,860 &  2.552\,745\,872 \\
    \dots  & \dots & \dots & \dots \\
  \end{tabular}
  \caption{The eigenvalue spectrum (for $l=2$) for the models I,II, and
   III. The non--stratified model~I has no g--modes.}
  \label{tab:1FluidEigenvalues}
\end{table}

\begin{figure*}
  \centering
  \includegraphics[width=17cm,clip]{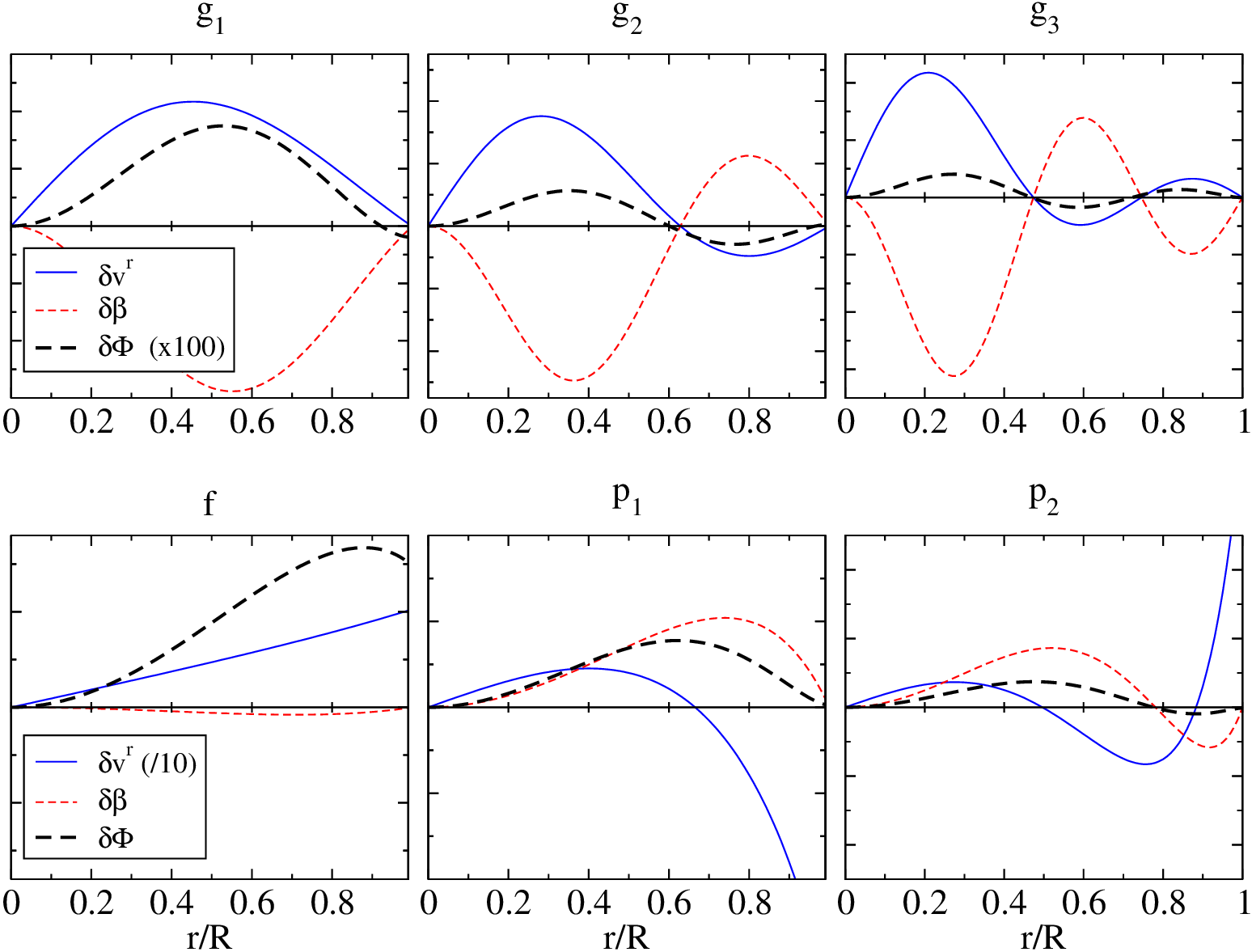}
  \caption{The first few ($l=2$) g--, f-- and p--modes for model
  II. We note that the g--modes are characterised by a very small
  radial velocity $\d v^r$ at the surface, while it is very large in
  the case of p--modes.}
\end{figure*}

We also see that in the non--stratified model~I, the one--fluid
frequencies and modes correspond exactly to the corresponding
``ordinary''-type solutions of the two--fluid case (see
table~\ref{tab:Eigenvalues2} and Fig.~\ref{fig:EigenmodesModelI}),
as would be expected from the separability of the system as discussed
in Sect.~\ref{sec:DecouplingOrdSf}.

We note that the perfect fluid modes of stratified models generally
have $\dbeta\not=0$, because adiabatic oscillations generally drive
fluid elements out of equilibrium, only in the non--stratified case
(model~I) is $\dbeta=0$ strictly satisfied. 

The absence/presence of $g$--modes in superfluid/normal fluid models
might seem somewhat surprising
but one can get a
better intuitive understanding by considering the physical origin of
these $g$--modes: a radially displaced fluid element will remain close
to mechanical (pressure) equilibrium with its surroundings, but its
respective values of $\rho_\n$ and $\rho_\c$ will generally differ
from the surroundings (i.e. when $\xc'\not=0$) and 
therefore (via the equation of state) its total density will also be
different, resulting in a buoyant restoring force and a corresponding
oscillation mode (in unstable models  this restoring force will
actually drive the fluid element still further away from its initial
position, leading to convection). 
In the simple (cold) superfluid models considered here,
each fluid element of either fluid ($\n$ or $\c$) is only
characterised by a \emph{single} quantity, namely $\rho_\n$ or
$\rho_\c$. Displacing an element of fluid $\n$, say,
will therefore result not only in mechanical equilibrium ($\mut^\n$),
but also in buoyant equilibrium. This can by seen by expressing its
density at the new position as $\rho_\n=\rho_\n(\mut^\n, \rho_\c)$.
The fluid $\c$ was not displaced, therefore not only $\mut^\n$ but
also $\rho_\c$ of the fluid element are identical to the background
values, and so is $\rho_\n$. 
If we had allowed for an additional comoving quantity like entropy
$s$, we would expect to find g--modes driven by a stratification in
$s/\rho_\n$. 

It is intriguing to see that the absence of the $g$--modes
in superfluid models is accompanied by an apparent doubling of
acoustic modes,  but it is not obvious to establish a link between
these different classes of modes as we are currently not aware of a
continuous transition from a two--fluid to a one--fluid model (either
the two fluids are locked together or they are not).

\section{Conclusions}
\label{sec:Conclusions}

In this paper we have tried to clarify the qualitative properties of
the eigenmode spectrum of superfluid neutron stars, using a simple
two--fluid model.
We have shown the important --- previously somewhat overlooked ---
role of stratification for these modes. The picture has been found to
be more complex than previous studies have suggested, and some of the
earlier conclusions have been shown to apply only for non--stratified
models. In particular, one can not generally talk about two distinct
families of ``superfluid'' and ``ordinary'' modes. The system of
equations describing two--fluid modes can not be separated in the case
of  stratified stars, and its solutions have no direct correspondence to
the eigenmodes of the one--fluid system. The two--fluid modes are
generally neither co-- nor counter-moving, rather \emph{all} of them
are characterised by non--zero amplitudes of relative velocity
$\d\vDv$, deviation of chemical equilibrium $\dbeta$ and total density
perturbation $\d\rho$. Also the order of the mode does not necessarily
correspond to the number of radial nodes (as seen in
Fig.~\ref{fig:p6Eigenmode}), which is possible because the system is
not of the Sturm--Liouville type.  
We have further confirmed earlier findings about the absence of
g--modes in these superfluid models
\citep{Lee95,andersson01:_dyn_superfl_ns}, as well as the appearance
of avoided crossings between mode frequencies when changing the
entrainment parameter \citep{andersson02:_oscil_GR_superfl_NS}. 

Given the radical difference and richer structure of the oscillations 
of superfluid neutron star models as  compared to the simple
perfect fluid models, we think that much future effort is needed to
further clarify these properties and evaluate possibly observable
consequences. The respective absence and presence  of g--modes in
these two different models is a striking example of such a
potentially observable indicator of superfluidity in neutron
stars. However, many more physical effects have to be taken into 
account in order to achieve a more realistic description of superfluid 
neutron stars, namely the inclusion of vortex--forces and beta
reactions, both of which will lead to dissipation. Furthermore, an
``envelope'' or an elastic crust should be included, and maybe most
importantly, the effects of rotation and magnetic field, which add
new restoring forces and result in a much richer spectrum of modes.
Eventually, for a realistic study of oscillations of superfluid
neutrons stars, one needs to work in a generally relativistic
framework, as pioneered by
\citet{comer99:_quasinorm_modes_GR_superfl_NS} and 
\citet{andersson02:_oscil_GR_superfl_NS}. 
This step is crucially important also for the assessment of the
gravitational radiation emitted by these modes, and their
stability/instability via the CFS mechanism \citep{friedman78:_secul_instab}.

\begin{acknowledgements}
We thank N.~Andersson, D.~Langlois, D.~Gondek, J.L.~Zdunik, H.~Beyer,
G.L.~Comer and B.~Dintrans for very valuable discussions in the early stages
of this work. We also thank D.I.~Jones and N.~Andersson for a careful
reading of the manuscript.

RP acknowledges support from the EU Programme 'Improving the Human
Research Potential and the Socio-Economic Knowledge Base' (Research
Training Network Contract HPRN-CT-2000-00137).

\end{acknowledgements}

\bibliographystyle{aa}
\bibliography{/home/rp/work/biblio}

\end{document}